\documentclass[letterpaper,11pt]{article}
\pdfoutput=1

\usepackage{jheppub} 

\usepackage{color}

\usepackage{verbatim}
\usepackage{amsmath}
\usepackage{slashed}

\usepackage{amssymb}
\usepackage[caption=false]{subfig}

\usepackage{multirow}
\usepackage{cleveref}

\newcommand{\be}{\begin{equation}}
\newcommand{\ee}{\end{equation}}

\newcommand{\gev}{\mathrm{GeV}}

\usepackage{color}
\definecolor{mit-red}{rgb}{0.64,.12,0.2}
\definecolor{darkred}{rgb}{1.0,0.1,0.1}
\definecolor{darkgreen}{rgb}{0.1,0.7,0.1}
\definecolor{darkblue}{rgb}{0.1,0.1,1.0}

\DeclareMathOperator\erf{erf}

\begin{document}

% Magic to fix footnote behavior.
\count\footins = 1000
\interfootnotelinepenalty=10000
\setlength{\footnotesep}{0.6\baselineskip}

%Change tables
\renewcommand{\arraystretch}{1.3}

\title{Discovery Prospects for a Minimal Dark Matter Model at Cosmic and Intensity Frontier Experiments}
\author{Ahmed Alenezi$^1$,}
\author{Cari Cesarotti,$^2$}
\author{Stefania Gori,$^3$}
\author{Jessie Shelton$^{1}$}

\affiliation[1]{Illinois Center for the Advanced Study of the Universe and Department of Physics, Grainger College of Engineering, University of Illinois at Urbana-Champaign, Urbana, IL 61801, USA}
\affiliation[2]{Center for Theoretical Physics, Massachusetts Institute of Technology, Cambridge, MA 02139, USA}
\affiliation[3]{Department of Physics and Santa Cruz Institute for Particle Physics, University of California Santa Cruz, Santa Cruz, CA 95064, USA}

\emailAdd{alenezi6@illinois.edu}
\emailAdd{ccesar@mit.edu}
\emailAdd{sgori@ucsc.edu}
\emailAdd{sheltonj@illinois.edu}
%\date{\today}

\preprint{MIT--CTP 5847}

%%%%%%%%%%%%%%%%%%%%%%%%%%%%%%%%%%%%%%%%%%%%%%%%%%%%%%%%%
\abstract{
We explore the detection prospects for a minimal secluded dark matter model, where a fermionic dark matter particle interacts with the Standard Model (SM) via a kinetically mixed dark photon. We focus on scenarios where the dark photon decays visibly, making it a prime target for beam-dump experiments. In this model, the dark matter relic abundance can be achieved by a variety of mechanisms: freeze-in, out-of-equilibrium freeze-out, and secluded freeze-out. We demonstrate that the secluded freeze-out regime in the considered mass range is now entirely excluded by a combination of direct and indirect detection constraints. Moreover, we show that future direct detection and intensity frontier experiments offer complementary sensitivity to this minimal model in the parameter space where the hidden sector never enters equilibrium with the SM.  In out-of-equilibrium freeze-out scenarios, nuclear-recoil direct detection experiments can still access signals above the neutrino fog that are mediated by dark photons that are too weakly coupled to be detected in future beam dump experiments.  Meanwhile, future beam dump experiments provide a powerful probe of the freeze-in parameter space in this model, which is largely inaccessible to direct detection experiments.  Notably, even in the absence of a future observation in direct detection experiments, a dark photon discovery remains possible at SHiP, DUNE, LHCb, and DarkQuest within this minimal dark matter model.
}
%%%%%%%%%%%%%%%%%%%%%%%%%%%%%%%%%%%%%%%%%%%%%%%%%%%%%%%%%
\maketitle

%%%%%%%%%%%%%%%%%%%%%%%%%%%%%%%%%%%%%%%%%%%%%%%%%%%%%%%%%
\section{Introduction}
\label{sec:intro}
%%%%%%%%%%%%%%%%%%%%%%%%%%%%%%%%%%%%%%%%%%%%%%%%%%%%%%%%%
%
Understanding the nature of dark matter (DM) is one of the most pressing challenges in particle physics.
Cosmological and astrophysical observations provide compelling evidence for its existence, yet its fundamental nature remains unknown.
The absence of a discovery of weakly interacting massive particle (WIMP) dark matter has strongly motivated the exploration of alternative dark matter candidates.
 In particular, scenarios featuring dark matter with a mass below the electroweak scale and residing in its own hidden sector have gained significant attention \cite{Abercrombie:2015wmb, Alexander:2016aln,Battaglieri:2017aum}.

A wide range of experimental efforts are currently underway to search for such candidates, including direct detection experiments probing dark matter scattering, accelerator-based experiments designed to produce and detect dark matter and dark-sector particles, and precise measurements of the cosmic microwave background (CMB).
For a recent overview of existing and planned searches, see the community studies \cite{Gori:2022vri,Antel:2023hkf}.

Dark photons are a possible minimal extension to the Standard Model (SM) that can provide a portal to a larger hidden sector~\cite{Ackerman:2008kmp, Feng:2008mu,Chu:2011be,Fabbrichesi:2020wbt}.
In this paper, we consider a minimal \emph{secluded} dark matter model, where dark matter is a Dirac fermion, $\chi$, charged only under a dark $U(1)$ gauge symmetry \cite{Pospelov:2007mp}.
 The dominant interaction between the dark matter and the SM is mediated by a kinetically mixed dark photon, $Z_D$, with mass $m_{Z_D} < m_{\chi}$.
 In this regime, the dark photon decays into SM particles, making it an important target for visibly decaying long-lived particle searches at beam-dump experiments, LHCb, Belle II, and dedicated LHC auxiliary detectors \cite{Antel:2023hkf}.

In this scenario, the dark matter relic abundance is determined by two key processes: direct dark matter production from the SM, $f\bar{f} \to \chi\bar{\chi}$, and dark matter annihilation into dark photons, $\chi\bar{\chi} \to Z_D Z_D$, which in turn decay back to the SM~\cite{Bernal:2015ova, Slatyer:2015jla, DAgnolo:2015ujb, DelNobile:2015uua, Evans:2017kti,Hambye:2019dwd,Fernandez:2021iti,Bhattiprolu:2023akk,Boddy:2024vgt}.
Above a certain value of the kinetic mixing parameter, $\epsilon$, the relic abundance curves become independent of $\epsilon$, and the dark matter abundance is controlled only by the dark gauge coupling, $\alpha_D$. 
For simplicity we refer to this value  of $\epsilon$ as the \textit{thermalization floor}. 
Below the thermalization floor, the dark sector temperature evolves non-adiabatically, requiring a full solution of coupled Boltzmann equations for the hidden sector temperature as well as the number of dark matter particles \cite{Chu:2011be}. 
 The observed dark matter density can be realized along a continuous curve in the $\alpha_D$-$\epsilon$ plane, which interpolates between a \emph{freeze-in} mechanism at small $\alpha_D$ and a \emph{secluded freeze-out} scenario at larger values of $\alpha_D$.

The aim of this work is to provide an up-to-date assessment of the detection prospects of this minimal model.
As we will demonstrate for the first time, the parameter space of this minimal model above the thermalization floor---the so-called \emph{WIMP-next-door} regime \cite{Evans:2017kti}---is now entirely excluded by a combination of direct and indirect detection constraints.
 However, large regions of parameter space of the model remain viable, particularly those where dark matter attains its relic abundance via the freeze-in mechanism or through freeze-out processes occurring while the hidden sector is undergoing non-adiabatic evolution.
In this paper, we also highlight exciting opportunities for future experiments:
 as we demonstrate in this work, a dark photon discovery at future beam-dump experiments such as DUNE \cite{Berryman:2019dme, DUNE:2020ypp}, SHiP \cite{SHiP:2021nfo, Alekhin:2015byh}, and DarkQuest \cite{Berlin:2018pwi,Apyan:2022tsd} would be fully consistent with this minimal dark matter scenario, even in the continued absence of signals in direct detection experiments. 
 Additionally, while much of the viable parameter space predicts dark matter-nuclear scattering cross-sections that lie below the so-called \emph{neutrino fog}, and is therefore challenging for direct detection experiments to probe, we find there still exist regions of parameter space where this model can be uniquely tested through future direct detection efforts. 

The paper is organized as follows:
In Sec.~\ref{sec:dmmodel}, we define the Lagrangian of the model and outline its physical parameter space, along with the thermal history of dark matter. 
In Sec.~\ref{sec:constraints}, we explore the experimental tests of this scenario, including CMB constraints, direct detection, and present and future accelerator-based searches.
 We reserve Sec.~\ref{sec:conclusions} for our conclusions. In Appendix~\ref{app:CrossSec}, we provide detailed expressions for the collision terms that enter the Boltzmann equation governing the cosmological evolution of the hidden sector.

%%%%%%%%%%%%%%%%%%%%%%%%%%%%%%%%%%%%%%%%%%%%%%%%%%%%%%%%%
\section{A Minimal Dark Matter Model}
\label{sec:dmmodel}
%%%%%%%%%%%%%%%%%%%%%%%%%%%%%%%%%%%%%%%%%%%%%%%%%%%%%%%%%

We begin by defining the minimal DM model that we will study in this paper.
 In Sec.~\ref{subsec:prelim} we write the Lagrangian of the model and establish notation. 
 Readers familiar with the extension of the SM by a kinetically-mixed U(1) gauge group can skip to Sec.~\ref{sec:boltz}, where we study the Boltzmann equations governing the evolution of the hidden sector particles in the early universe. 
 The numerical solution of these equations is presented in Sec.~\ref{sec:numsol}.

\subsection{Preliminaries}
\label{subsec:prelim}
 Let us review the minimal dark matter model which we consider in this paper: a Dirac fermion, $\chi$, charged under an additional U(1)$_D$ gauge group with a new gauge boson, $\hat{V}_\mu$.
\subsubsection*{Physical gauge bosons}
The gauge symmetry of the model is SU(2)$_L$ $\times$ U(1)$_Y$ $\times$ U(1)$_D$. 
Mixing occurs between the field strength terms of the two U(1) symmetries,
\begin{equation}
\label{eq:origLagKin}
\mathcal{L}_\text{kin} \supset -\frac{1}{4}\hat{V}_{\mu \nu} \hat{V}^{\mu \nu} -\frac{1}{4}\hat{B}_{\mu \nu} \hat{B}^{\mu \nu} + \frac{\epsilon}{2}\hat{V}_{\mu\nu} \hat{B}^{\mu \nu},
\end{equation}
where $\hat B$ is the gauge boson associated to the SM $U(1)_Y$. 
To diagonalize these kinetic terms, we move to the following basis:
\begin{equation}
\label{eq:firstBasis}
\begin{pmatrix}
X \\ B 
\end{pmatrix} 
= 
\begin{pmatrix}
\sqrt{1-\epsilon^2} & 0 \\
- \epsilon & 1
\end{pmatrix}
\begin{pmatrix}
\hat{V} \\ \hat{B}
\end{pmatrix}.
\end{equation}

A mass term for $\hat V$ can be added to the Lagrangian.
This term can come either from a Higgs or a St\"uckelberg mechanism \cite{Stueckelberg:1938hvi, Redi:2022zkt}. 
For minimality, we will consider a St\"uckelberg mass, $\mathcal L_m=\frac{1}{2}m_V^2\hat V^\mu\hat V_\mu$, though the resulting phenomenology is also broadly consistent with a dark Higgs mechanism.\footnote{We briefly comment on the phenomenology of the dark Higgs mechanism scenario in Sec.~\ref{sec:conclusions}.} 
The $\hat V$, $\hat B$, and gauge bosons of the SU(2)$_L$ group, $W^a$, further mix after adding an explicit mass term and after the SM electroweak symmetry breaking. 
Diagonalizing these mass terms yield the physical degrees of freedom that propagate as the various vector bosons of our theory.
The mass mixing of the neutral gauge bosons is
\begin{equation}
\mathcal{L}_\text{mass} \supset \frac{m_{Z_0}^2}{2} \begin{pmatrix} Z_0^\mu & X^\mu \end{pmatrix} 
\begin{pmatrix}
 1 & -\eta \sin\theta_W \\
 -\eta \sin\theta_W & \delta^2 + \eta^2 \sin^2 \theta_W
\end{pmatrix}
\begin{pmatrix} {Z_0}_\mu \\ X_\mu \end{pmatrix},
\label{eq:massMat}
\end{equation}
where we have defined $\delta \equiv \frac{m_V}{m_{Z_0}}$ and $\eta \equiv {\epsilon}/{\sqrt{1-\epsilon^2}}$ and $Z_0^\mu$ and $m_{Z_0}$ are the SM $Z$ boson and its mass before mixing.
The $W^\pm$ boson of the SM remain spectators to the reshuffling of degrees of freedom.

We diagonalize the mass Lagrangian in Eq.~\ref{eq:massMat} by the rotation
\begin{equation}
\begin{pmatrix} Z \\ Z_D \end{pmatrix} = \begin{pmatrix} \cos\alpha & \sin \alpha \\ -\sin\alpha & \cos \alpha \end{pmatrix}
\begin{pmatrix} Z_0 \\ X \end{pmatrix} 
\end{equation}
where
\begin{equation}
\label{eq:finalRot}
\tan 2\alpha = \frac{2 \eta \sin\theta_W}{1-\eta^2 \sin^2\theta_W - \delta^2}.
\end{equation}
The three mass eigenstates are then given by
\begin{equation}
\label{eq:dofm}
\begin{aligned}
&A_\mu = \cos\theta_W B_\mu + \sin\theta_W W_\mu^3 \\
& Z_\mu = \cos\alpha\left( \cos\theta_W W^3_\mu - \sin\theta_W B_\mu \right) \cos\alpha+ \sin\alpha X_\mu \\
&Z_{D\mu} = \cos\alpha X_\mu - \sin\alpha\left(\cos\theta_W W_\mu^3 + \sin \theta_W B_\mu \right).
\end{aligned}
\end{equation}
The photon $A_\mu$ remains massless, and the $Z_\mu$, $Z_{D\mu}$ have the masses
\begin{equation}
m_{Z,Z_D}^2 = \frac{m_{Z,0}^2}{2}\left( 1 + \delta^2 + \eta^2 \sin^2\theta_W \pm \text{Sign}(1-\delta^2) \sqrt{(1+\delta^2 + \eta^2 \sin^2 \theta)^2 - 4 \delta^2} \right).
\label{eq:massEig}
\end{equation}

\subsubsection*{Fermion-gauge boson interactions}
Fermions in both the SM and the hidden sector obtain couplings to both the $Z$ and $Z_D$ from the covariant derivative. 
Before the diagonalization to gauge boson mass eigenstates, the covariant derivative for the various fermions is
\begin{equation}
\label{eq:cov2}
\begin{aligned}
D_\mu \equiv &~ \partial_\mu - i \hat{g}_X Q_X \hat{V}_\mu - i g_Y Q_Y \hat{B}_\mu - i g T^a W^a_\mu \\
& \rightarrow \partial_\mu - i g_X Q_X X_\mu - i g_Y Q_Y \left( \eta X_\mu + B_\mu \right) - i g T^a W^a_\mu,
\end{aligned}
\end{equation}
 with $\hat g_X$, $g_Y$, and $g$ the couplings corresponding to the $U(1)_D$, $U(1)_Y$, and $SU(2)$ gauge symmetries, respectively, and $g_X\equiv \hat g_X/\sqrt{1-\epsilon^2}$.
To compute the interactions with the physical gauge bosons, we plug the new basis from Eq.~\ref{eq:dofm} into Eq.~\ref{eq:cov2} and find
\begin{eqnarray}\label{eq:cov3}
D_\mu & =& \partial_\mu -i g \sin\theta_W (Q_Y+T^3) A_\mu  \\\nonumber
&& - i\left( \frac{g}{\cos\theta_W}\left(\cos\alpha( T^3\cos^2 \theta_W - Q_Y\sin^2\theta_W) + \eta Q_Y \sin\alpha \sin \theta_W \right)+ g_X Q_X\sin\alpha \right)Z_\mu \\\nonumber
&& + i \left( \frac{g}{\cos\theta_W}\left(\sin \alpha (T^3 \cos^2 \theta_W-Q_Y \sin^2 \theta_W) - \eta Q_Y \cos\alpha \sin\theta_W\right)-g_X Q_X\cos \alpha\right) Z_{D \mu}.
\end{eqnarray}

The $T^3$ operator corresponds to the SU(2)$_L$ gauge symmetry, whereas the $Q$ is the electromagnetic charge after spontaneous symmetry breaking.
Without loss of generality, we can set $Q_X = 1$ for DM, while the SM fermions are not charged under $U(1)_D$ and therefore the corresponding $Q_X=0$

For ease of reading, we summarize the couplings in Table~\ref{tab:couplings}.

\begin{table}
\begin{center}
\begin{tabular}{ | c | c | c |}
\hline
Particle & $\bar{f}f$ & $\bar{\chi}\chi$ \\ 
\hline
 $Z$ & $\frac{g}{\cos\theta_W} \left( \cos\alpha (T^3 - Q \sin^2 \theta_W) + \eta \sin\alpha \sin\theta_W (Q-T^3)\right)$ & $\sin\alpha g_X$ \\ 
 \hline
$Z_D$ & $\frac{g}{\cos\theta_W} \left( \sin\alpha (Q \sin^2 \theta_W - T^3) + \eta \cos\alpha \sin\theta_W (Q-T^3)\right)$ & $\cos\alpha g_X$ \\ 
\hline
$A$ & $g \sin\theta_W Q$ & $0$ \\
\hline
\end{tabular}
\caption{Couplings of the gauge boson mass eigenstates with SM fermions and dark matter.}
\label{tab:couplings}
\end{center}
\end{table}

\subsubsection*{Summary of parameters}
There are four parameters that determine the phenomenology of our model. 
\begin{enumerate}
\item \textbf{Kinetic mixing parameter \boldmath $\epsilon$.} 
This parameter shows up in Table~\ref{tab:couplings} through $\eta$ and the angle $ \alpha$. We see that as $\epsilon \rightarrow 0$, $\eta \rightarrow \epsilon$ and subsequently $\alpha, \eta \rightarrow 0$ from Eq.~\ref{eq:finalRot}.
 Thus, as the mixing weakens, the interactions between the two sectors turns off. 
\item \textbf{Hidden sector coupling \boldmath $\alpha_D$.} We define this to be the coupling of the dark photon $Z_D$ to the DM, such that $\alpha_D \equiv \frac{ g_X^2}{4\pi}$. 
\item \textbf{Dark photon mass \boldmath $m_{Z_D}$}. We can choose the mass before SM spontaneous symmetry breaking, but this value will get a correction after mass mixing with the $Z$.
 Its dependence is shown in Eq.~\ref{eq:massMat}, but since the original mass $m_V$ is a free parameter, we can scan over values of $m_{Z_D}$ independently of $\epsilon$. 
\item \textbf{Dark matter mass \boldmath $m_\chi$}. This is a parameter that we can set. In this work we choose to always be in the regime $m_\chi > m_{Z_D}$ such that $Z_D$ decays visibly. 
\end{enumerate}
The dimensionality of the parameter space of our model is reduced by one degree of freedom after imposing the observed relic abundance, which we discuss in Sec.~\ref{sec:numsol}.
Throughout this work we will explore which experiments are sensitive to certain regimes of the parameters as well as the resultant phenomenology. 

%%%%%%%%%%%%%%%%%%%%%%%%%%%
\subsection{Thermal History}
\label{sec:boltz}
%%%%%%%%%%%%%%%%%%%%%%%%%%%
%

\begin{figure} \centering
\includegraphics[width=0.4\textwidth]{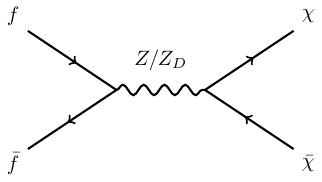} \qquad
\includegraphics[width=0.305\textwidth]{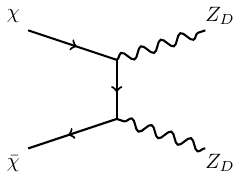}\qquad
\caption{The relevant processes for achieving the measured relic abundance. SM fermion annihilations to DM pairs transfer both energy and DM abundance into the the hidden sector via $f\bar f \rightarrow \chi \bar\chi$ (left), while DM annihilations to dark photons within the hidden sector $\chi\bar\chi \rightarrow Z_D Z_D$ (right) deplete the DM abundance.} 
\label{fig:feynman}
\end{figure}

The primary processes governing the evolution of our DM model in the early universe are $f\bar{f} \leftrightarrow \chi\bar\chi$, and $\chi\bar\chi \leftrightarrow Z_D Z_D$ (see Fig.~\ref{fig:feynman}).
The first of these processes, where SM fermions annihilate into DM pairs, dominates the energy transfer between the SM and the hidden sector (HS).
 When the HS is weakly enough coupled to the SM to be out of equilibrium in the early universe, this interaction is responsible for the non-adiabatic evolution of the HS temperature as well as contributing directly to the DM abundance.
The second process, DM annihilation into dark photons and its inverse, can also play an important role in determining the DM abundance for sufficiently large values of $\alpha_D$.
We focus on DM masses below 100 GeV, for which DM production is dominated by temperatures below electroweak symmetry breaking. 
We consider the parameter space where $m_{Z_D} < m_\chi$ such that $Z_D$ can only decay visibly, and in particular consider the regime with $m_{Z_D} \lesssim m_\chi/10$. 
In this part of parameter space, the dark photon is still approximately relativistic when DM annihilations decouple.
 This simplifies the thermal history of the model considerably, for three distinct reasons.
 First, the cross-sections describing DM annihilation become independent of $m_{Z_D}$ in this regime, which helps streamline the parameter space of the model. 
Second, the dark photon can be approximated as a relativistic particle in thermal equilibrium at the hidden sector temperature throughout the DM freeze-out process, so that the dark photon number density does not need to be separately tracked in the Boltzmann equations.
Third, requiring $m_{Z_D}\lesssim \tilde T_\mathrm{fo}$, the temperature of the hidden sector at freeze-out, ensures that neglecting the energy transfer into the HS from direct dark photon production is a reasonable approximation in evaluating the relic abundance of dark matter in the regime where the HS does not thermalize with the SM. 

In particular, at SM temperatures $T\gg m_{Z_D}$ we can neglect direct dark photon production from the SM plasma since it is suppressed by in-medium corrections to the effective mixing angle \cite{An:2013yfc,Redondo:2013lna}, in contrast to the annihilations of SM fermions to DM pairs, which proceed without this suppression \cite{Knapen:2017xzo}.
 As pointed out in \cite{Hardy:2016kme}, this suppression can be understood as a generic consequence of having a particle whose coupling to the particles constituting a given thermal medium is generated entirely by mixing with a single SM degree of freedom within that medium. 
The dark photon mixes with the hypercharge gauge boson and thus does not couple exactly like a photon, i.e., for a massive dark photon it is not obvious that we can assume the $Z_D$ couplings to the SM take the simple photon-like form realized in the limit $\delta\to 0$, where this in-medium cancellation holds.
 However, we checked that corrections to the thermal dark photon production rate at finite values of $\delta$ are $\mathcal{O}(\delta^4)$, which ensures that this rate remains negligible in the mass range of interest to us; see also \cite{Heeba:2019jho}. 
 Thus taking $m_{Z_D} \ll m_\chi$ (and $m_{Z_D} \ll m_Z$) ensures that direct dark photon emission from the SM does not have a significant impact on the temperature evolution of the hidden sector while the DM abundance is evolving.

We now turn to the collision terms governing the thermal history of DM. 
In evaluating the collision terms we employ Maxwell-Boltzmann statistics, which is a good approximation since the relevant DM production processes are IR-dominated. 
This choice also significantly streamlines the calculations.
 Each scattering process can be expressed in terms of a Lorentz-invariant amplitude square function $\mathcal{A} (s)$ of the Mandelstam variable $s$:
\begin{equation}
\mathcal{A} (s)\equiv \int d\Pi_f (2\pi)^4 \delta^4 (\sum p) \left| \overline{\mathcal{M}} \right|^2 .
\label{eq:As}
\end{equation}
Here the matrix element $ \left| \overline{\mathcal{M}} \right|^2 $ is averaged over initial state spin and summed over final state spins.
In App.~\ref{app:CrossSec}, we present detailed expressions for $A(s)$ for both processes shown in Fig.~\ref{fig:feynman}.

Both the $Z$ and $Z_D$ can mediate the annihilation process $f \bar f \rightarrow \chi \bar\chi$, where $f$ is any SM fermion $f$.
 We regulate the $Z$ propagator with its (zero-temperature) decay width and include contributions from the resonant region in evaluating the collision integrals. 
 Retaining the resonant region is equivalent to including the contribution of on-shell $Z\to \chi\bar\chi$ decays to DM pair production \cite{Giudice:2003jh}. 
This annihilation process contributes to the forward collision term for DM number density (per Ref.~\cite{Gondolo:1990dk}) as
\begin{equation}
\begin{aligned}
\mathcal{C}^n_{f\bar f \rightarrow \chi \bar \chi} (T) \equiv &\, n_\text{eq}(T)^2 \langle \sigma v \rangle_{f\bar f \rightarrow \chi \bar \chi} \\
= & \int d \Pi_i |\overline{\mathcal{M}}_{f\bar f \rightarrow \chi \bar \chi}|^2 (2\pi)^4 \delta^{4}(\sum_i p_i) f_f^\text{eq}(T) f_{\bar{f}}^\text{eq}(T) \\
= & \frac{g_fg_{\bar{f}}}{8\pi^4} \frac{T}{8} \int _{\max(4m_{\chi},4 m_f^2)}^\infty ds \sqrt{s-4m_f^2} \mathcal{A}_{f\bar f \rightarrow \chi \bar \chi} (s) K_1(\sqrt{s}/T),
\end{aligned}
\label{eq:colTermN_1}
\end{equation}
where $g_f$, $g_{\bar{f}} =2$ count the internal degrees of freedom of the fermions and anti-fermions in the initial state. 
In this equation, $K_1(u)$ is the modified Bessel function of the first kind and we assume the SM particles $f$, $\bar f$ are in thermal equilibrium at a temperature $T$ that follows a Maxwell-Boltzmann distribution with number density $n_\text{eq}(T)$. 

Similarly, the forward collision term for the energy density following from fermion annihilation is 
\begin{equation}
\begin{aligned}
\mathcal{C}^\rho_{f\bar f \rightarrow \chi \bar \chi} (T) \equiv & \,n_\text{eq}(T)^2 \langle \sigma v E \rangle_{f\bar f \rightarrow \chi \bar \chi} \\
 = & \int d \Pi_i |\overline{\mathcal{M}}_{f\bar f \rightarrow \chi \bar \chi}|^2 (2\pi)^4 \delta^{4}(\sum_i p_i) f_f^\text{eq}(T) f_{\bar{f}}^\text{eq}(T) (E_f + E_{\bar f}) \\
= & \frac{g_fg_{\bar{f}}}{8\pi^4} \frac{T}{8} \int _{\max(4m_{\chi},4 m_f^2)}^\infty ds \sqrt{s}\sqrt{s-4m_f^2} \mathcal{A}_{f\bar{f}\rightarrow\chi\bar\chi}(s) K_2(\sqrt{s}/T).
\end{aligned}
\end{equation}
The backward collision terms $\mathcal{C}^n_{\chi \bar \chi \rightarrow f \bar f}$ and $\mathcal{C}^\rho_{\chi \bar \chi \rightarrow f \bar f}$ are similarly obtained, and can be found from the above expressions by taking $T\rightarrow \tilde T$, $m_{f}\leftrightarrow m_{\chi}$ and $n_{eq}\rightarrow n_\chi$.

DM annihilation to dark photons and the reverse process also contribute to the collision term for the DM number density. 
When the DM is in kinetic equilibrium at the HS temperature $\tilde T$, the corresponding contribution to the forward collision term $C^n$ can be written as 
\begin{equation}
\begin{aligned}
\mathcal{C}^n_{\chi \bar \chi\to Z_D Z_D} (\tilde T)  = &\,n_{\mathrm{eq}}^2 (\tilde T)
 \langle \sigma v \rangle_{ \chi \bar \chi \to  Z_D Z_D } \\
 = & \frac{g_\chi g_{\bar{\chi} }}{8\pi^4} \frac{\tilde T}{8} \int _{max(4m_{\chi},4 m_{Z_D}^2)}^\infty ds \sqrt{s-4m_\chi^2} \mathcal{A}_{\chi \bar \chi \to Z_D Z_D} (s) K_1(\sqrt{s}/\tilde T) 
\end{aligned}
\label{eq:colTermN_2}
\end{equation}
The amplitude square function for this process is also given in App.~\ref{app:CrossSec}.
We take $m_{Z_D}$ to zero in evaluating $C^n$, as the DM annihilation cross-section to dark photons becomes independent of $m_{Z_D}$ in the regime $m_\chi \geq 10 ~m_{Z_D}$.
 The dark photon mass will become important when we discuss direct and indirect detection in the following section. 
 The assumption that the HS is at internal kinetic equilibrium at $\tilde T$ throughout the freeze-out process is self-consistent for the DM masses and couplings $\alpha_D$ of interest to us here \cite{Fernandez:2021iti}.

The Boltzmann equations describing the HS temperature and the DM abundance are 
\begin{equation}
\begin{aligned}
\dot{\rho}_{HS}+3H(\rho_{HS}+P_{HS}) &= C^{\rho} \\
\dot{n}_{\chi}+3Hn_{\chi}&= C^n ,
\end{aligned}
\label{eq:boltzsystem}
\end{equation}
where a dot indicates a derivative with respect to time, and the collision terms $C^\rho$, $C^n$ include backward as well as forward processes.
We make two simplifying approximations: first, that the Hubble rate is dominated by the SM, so that we can take $H \approx H(T)$, and second, that the SM temperature evolves adiabatically. 
These are good approximations for small ($g_{*,HS} \ll g_{*,SM}$) and/or cold ($\tilde T\ll T$) hidden sectors, which will be our areas of interest. 
These approximations determine the evolution of the SM temperature $T(a)$ and the Hubble rate $H$, so that the only coupled equations we need to solve numerically are the two in Eq.~\ref{eq:boltzsystem}.

In solving these equations, we write $\rho_{HS}=\rho_{Z_D}+2\rho_\chi$ where the factor of 2 accounts for the contributions of $\chi$ and $\bar \chi$. Recalling that our DM is Dirac, we will ultimately require $\rho_\chi$ to make up half the observed DM relic abundance.
We take the energy density and pressure of the $Z_D$ to be given by the equilibrium distribution for a relativistic boson, $\rho_{Z_D}= 3 P_{Z_D} = g_{Z_D}(\frac{\pi^2}{30})\tilde T^4$.
 For the dark matter, in order to incorporate the contribution from its rest energy when non-relativistic, we adopt the Maxwell-Boltzmann results 
\begin{equation}
\begin{aligned}
\rho_\chi &= \left(m_\chi \frac{K_1(m_\chi /\tilde T)}{K_2(m_\chi /\tilde T)} + 3 \tilde T\right) n_\chi \equiv B(\tilde{T}) n_\chi \\
P_\chi&= \tilde T n_\chi .
\end{aligned}
\label{eq:DMrho}
\end{equation}

We use $x\equiv m_\chi/T$ as our independent variable and rewrite Eq.~\ref{eq:boltzsystem} to solve for $\tilde T$ and the DM yield $Y_\chi \equiv n_\chi / s(T)$, where $s(T) =\frac{2\pi^2}{45} g_{*S, SM} T^3$ is the SM entropy density. 
Using the conservation of SM entropy, we translate a time derivative into a derivative with respect to $x$ as
\begin{equation}
\begin{aligned}
\frac{d}{dt} = \left(\frac{Hx}{1+\frac{T}{3g_{*S}} \frac{dg_{*S}}{dT}}\right) \frac{d}{dx} \equiv A \frac{d}{dx} 
\end{aligned}
\label{eq:ddt_to_ddx}
\end{equation}
The coupled Boltzmann equations we solve numerically are then: 

\begin{equation}
\begin{aligned}
\frac{d\tilde T}{dx} &= \Bigg(s^2\sum_f \langle\sigma v E \rangle _{f\bar f \rightarrow \chi \bar \chi}Y_f^2(T) -s^2\sum_f \langle\sigma v E\rangle_{\chi\bar \chi \rightarrow f \bar f}Y_\chi^2-4H\rho_{Z_D}-2 sA B\frac{dY_\chi}{dx} - 6 H s Y_\chi \tilde{T}\Bigg) \times  \\
& \left(A\frac{d \rho_{Z_D}}{d \tilde{T}} + 2 sY_\chi A\frac{dB}{d\tilde{T}} \right)^{-1}\\
\frac{dY_{\chi}}{dx}& =\frac{s}{A} \left(-{\langle\sigma v\rangle_{\chi \bar \chi \to Z_D Z_D}}\left( Y_{\chi}^2-Y_{\rm eq}^2(\tilde T) \right)+ \sum_f \langle\sigma v\rangle_{f\bar f \rightarrow \chi \bar \chi}Y_f^2(T)-\sum_f \langle\sigma v\rangle_{\chi\bar \chi \rightarrow f \bar f}Y_\chi^2\right).
\end{aligned}
\label{eq:boltz_final}
\end{equation}

For each DM mass, the DM can attain its observed relic abundance along a curve in the two-dimensional space parameterized by $\alpha_D$ and $\epsilon$ \cite{Chu:2011be}. 
We next solve these equations numerically.
For all points considered, the lifetime of the dark photon is sufficiently short to avoid constraints from BBN ($\tau_{Z_D} \ll 1$ s).

%%%%%%%%%%%%%%%%%%%%%%%%%%%%%
\subsection{Numerical solutions}
\label{sec:numsol}
%%%%%%%%%%%%%%%%%%%%%%%%%%%%%

In this section, we present the numerical solutions to the Boltzmann equations of Eq.~\ref{eq:boltz_final}. 
To solve these equations, we start the numerical evolution at $\tilde{x}_0=m_\chi/\tilde T_0 = 10^{-2}$ with the initial HS temperature $ \tilde T_0 = 10^{-4} T_0$, and take the initial DM yield $Y_{\chi,0}=0$.
The values of the initial SM temperature $T_0$ are chosen as a function of DM mass such that the two sectors have ample time to interact before the final DM relic abundance is determined. 
We choose a small initial value of $\tilde T_0$ to ensure that the HS remains sufficiently cold to recover the freeze-in regime for small values of $\epsilon^2\alpha_D$---i.e., to ensure that DM production from the SM dominates over any DM population from the initial conditions.
For values of $\alpha_D$ and $\epsilon$ that yield solutions with larger HS temperatures than the initial condition, the HS temperature rapidly rises to the correct solution~\cite{Evans:2019vxr,Fernandez:2021iti}.

\begin{figure}[h!]
 \centering
 \includegraphics[width=0.8\linewidth]{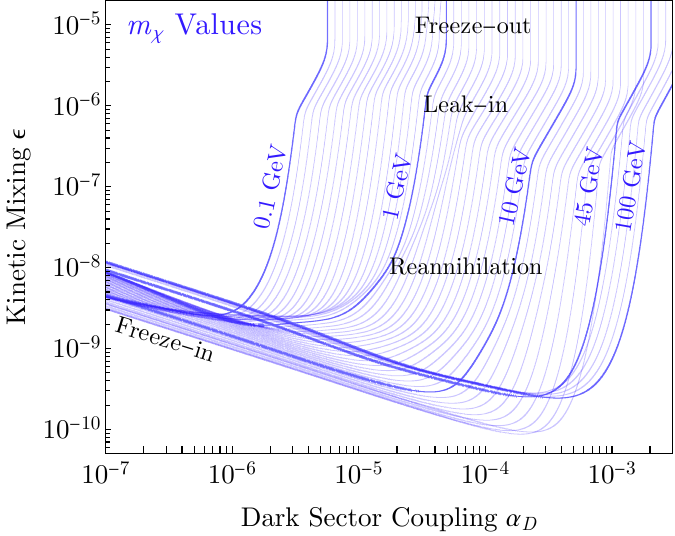}
 \caption{Contours in the $\epsilon, \alpha_D$ plane that realize the measured relic abundance for a given value of $m_\chi$. %s
 In this plot we see four distinct regions, which we have labeled: freeze-in, reannihilation, leak-in, and freeze-out. 
 We refer to the the reannihilation and leak-in phases collectively as out-of-equilibrium freeze-out.
 The regions are broadly indicated, with the boundaries between them occurring at significant changes in the slope of the relic abundance contours for each value of the DM mass.
 Note that at the final transition from leak-in to (secluded) freeze-out, the relic abundance condition becomes independent of the kinetic mixing parameter, $\epsilon$.
 }
\label{fig:Relic_Abundance}
\end{figure}

We plot the contours of $\alpha_D$ and $\epsilon$ that yield the measured relic abundance for DM masses $m_\chi =100 $ MeV to 100 GeV in Fig.~\ref{fig:Relic_Abundance}. Here we require $\Omega h^2 =0.1186\pm 0.0020$ \cite{Planck:2018vyg, Dvorkin:2022bsc}. 
%
%%%%%%%%%%%%%%%%%%%%%%
%%%%%%%%%%%%%%%%%%%%%%%%%
%
There are four distinct regions along each mass contour. 
At small $\alpha_D$, the curves start out in the ``freeze-in'' regime such that DM abundance is determined only by direct DM production from the SM, which is proportional to $\epsilon^2 \alpha_D$. 
As $\alpha_D$ becomes larger, the process $\chi \bar \chi \to Z_D Z_D$ starts to appreciably deplete DM, which means that more DM (and thus more energy) must be provided by out-of-equilibrium production from the SM to obtain the right DM abundance.
 In this regime, the curves in Fig.~\ref{fig:Relic_Abundance} require $\epsilon$ to increase with increasing $\alpha_D$. 
 The dynamics in this regime result from a nontrivial interplay of out-of-equilibrium processes with equilibrium processes within the hidden sector, giving rise to steep ``reannihilation'' \cite{Chu:2011be,Berger:2018xyd} and more shallowly sloped ``leak-in'' \cite{Evans:2019vxr} phases.\footnote{The characteristic out-of-equilibrium dynamics realized here rely on the existence of a dark radiation bath, which is the key motivation for our restriction $m_{Z_D}< m_\chi/10$. Adding an additional dark radiation species to furnish a ``dark sink''  \cite{Bhattiprolu:2023akk} allows these out-of-equilibrium dynamics to be realized for less-restrictive choices of the dark photon mass.} 
We refer to these two phases collectively as ``out-of-equilibrium freeze-out''. 
Finally, at sufficiently large values of $\epsilon$, the HS equilibrates with the SM before DM freeze-out. In this ``WIMP-next-door'' regime, the DM relic abundance becomes independent of $\epsilon$. 
The minimum value of $\epsilon$ that allows for this equilibration, as a function of $m_\chi$, defines a \textit{thermalization floor}\footnote{These results update and correct the vector portal results of \cite{Evans:2017kti}, which did not account for the in-medium suppression of the kinetic mixing in evaluating the thermalization floor.}. 

It will be convenient in what follows to split up the predictions of this model into three distinct regimes: (i) freeze-in, (ii) out-of-equilibrium freeze-out, and (iii) above the thermalization floor. 
Quantitatively, we identify the thermalization floor $\alpha_D^{TF}$ for each DM mass by finding the value of $\alpha_D$ at which the relic abundance curve becomes $\epsilon$-independent.  It is worth bearing in mind a subtlety here: for values of $\epsilon$ immediately above this ``thermalization floor,'' the dark sector is not actually equilibrated with the SM throughout the freeze-out process.  Rather, in this regime the energy injection from the SM into the dark sector after kinetic decoupling of the two sectors is negligible in comparison with the red-shifting of the HS energy bath, so that the HS temperature still evolves adiabatically.

We separate the freeze-in regime from the out-of-equilibrium freeze-out regime using the minimum value of $\epsilon$ on the relic curve in Fig. \ref{fig:Relic_Abundance}, so that the freeze-in branch refers to the portion of the relic curve with $\alpha_D < \alpha_D(\epsilon_\mathrm{min})$. 
 Thus the out-of-equilibrium freeze-out regime is characterized by $\alpha_D(\epsilon_\mathrm{min}) <\alpha_D < \alpha_D^{TF}$.\footnote{Dark photons with $\epsilon < \epsilon_\mathrm{min}$ can still be consistent with this minimal model of DM if the hidden sector is populated by some non-SM source in the early universe \cite{Arvanitaki:2021qlj,Fernandez:2021iti}. Here we focus on the more predictive scenario where interactions with the SM dominate the production.}

%%%%%%%%%%%%%%%%%%%%%%%%%%%%%%%
\begin{figure}[t!]
  \centering
  \includegraphics[width=0.45\linewidth]{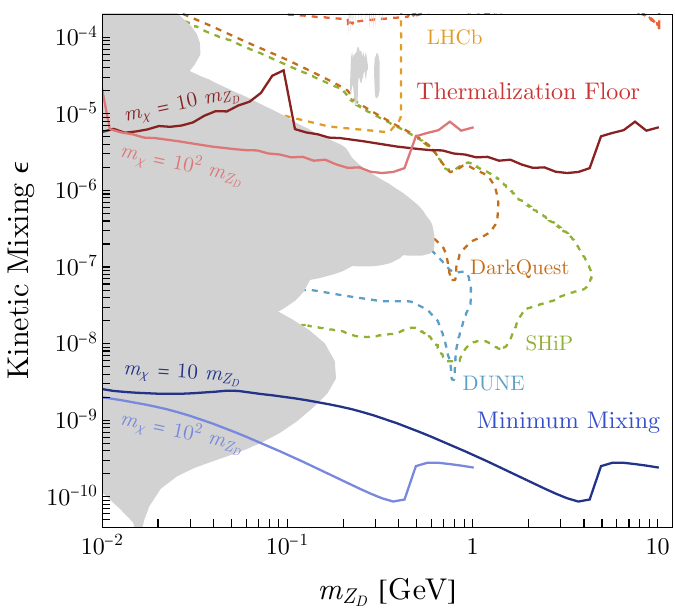} 
  \caption{The thermalization floor (red) and minimum mixing $\epsilon_\text{min}$ separating freeze-in and freeze-out regimes (blue) mapped into the $m_{Z_D}$-$\epsilon$ plane, for fixed $m_{\chi}/m_{Z_D} = 10$ (darker) and $m_{\chi}/m_{Z_D} = 100$ (lighter). Additional curves showing excluded regions (shaded gray) and projected reach (colored dotted lines) from accelerator-based experiments are taken from Refs.~\cite{Batell:2022dpx, SHiP:2021nfo}, while exclusions from Supernova 1987a are taken from from Ref.~\cite{Chang:2016ntp}.}
 \label{fig:TFFIcontours}
\end{figure}
%%%%%%%%%%%%%%%%%%%%%%%%%%%%%%%

In Fig.~\ref{fig:TFFIcontours}, we show how the thermalization floor and the minimum-mixing line $\epsilon_\mathrm{min}$ can be translated into the dark photon parameter space.
The thermalization floor and minimum-mixing line are defined as a function of DM mass.
To map them into the dark photon parameter space, we consider fixed values of the mass ratio $m_{Z_D}/m_\chi$. 
For illustrative purposes, we show results for $m_\chi = 10 m_{Z_D}$ and  $m_\chi = 100 m_{Z_D}$. As we can see from Fig.~\ref{fig:Relic_Abundance} and Fig.~\ref{fig:TFFIcontours}, both the QCD phase transition and DM production from $Z$ decays give the thermalization floor and the minimum-mixing a notable $m_\chi$ dependance, in contrast to the simpler $m_\chi$ dependence exhibited away from mass thresholds \cite{Evans:2019vxr,Fernandez:2021iti}. 
In Sec.~\ref{sec:constraints} we will discuss in detail the interplay between searches for the dark photon and constraints on DM from direct and indirect detection.

%%%%%%%%%%%%%%%%%%%%%%%%%%%%%%%%%%%%%%%%%%%%%%%%%%%%%%%%%
\section{Experimental Tests}
\label{sec:constraints}
%%%%%%%%%%%%%%%%%%%%%%%%%%%%%%%%%%%%%%%%%%%%%%%%%%%%%%%%%

This simple dark matter model has a velocity- and spin-independent cross-section with nuclei and an $s$-wave annihilation cross-section, and it is therefore subject to stringent constraints over much of its parameter space from both direct and indirect detection experiments.
In this section we delineate these constraints, in particular evaluating limits on this model in the region where the hidden sector is out of thermal equilibrium with the SM. 
In doing so we demonstrate that there is viable and complementary discovery prospects for this minimal DM model in both nuclear-recoil direct detection (DD) experiments as well as in accelerator based experiment searches for visibly-decaying dark photons.

%%%%%%%%%%%%%%%%%%%%%%%%%%%%%%%%%%%%%%%%%%%%%%%%%%%%%%%%%
\subsection{Indirect Detection}
\label{sec:IndDet}
%%%%%%%%%%%%%%%%%%%%%%%%%%%%%%%%%%%%%%%%%%%%%%%%%%%%%%%%%
%
Since dark photons decay promptly into pairs of SM particles, the products of DM annihilations are energetic SM particles, which can leave imprints in astrophysical and cosmological datasets.
We focus here on the constraints on DM annihilation provided by the anisotropy spectrum of the cosmic microwave background (CMB). 
Fermi-LAT observations of dwarf galaxies \cite{Fermi-LAT:2015att,Fermi-LAT:2016uux} as well as AMS-02 observations of positron spectra \cite{AMS:2013fma} can provide similar constraints on the DM annihilation cross-section and can nominally contribute additional sensitivity for DM masses above $10$ GeV.
However, searches for DM annihilation products from galaxies today are subject to sizable systematic uncertainties relating to the modeling of DM haloes and astrophysical sources of cosmic rays. 
CMB constraints, on the other hand, primarily depend on the properties of the universe at times when it was well-described by perturbation theory.   In addition, as shown for this model in \cite{Evans:2017kti}, the overall sensitivity of cosmic ray searches is broadly comparable to that from the CMB. For this reason, we do not add cosmic ray observations to our list of constraints.

The parameter space of our model of dark matter is constrained by the CMB anisotropies as measured from Planck \cite{Planck:2015fie}, which restricts \cite{Slatyer:2015jla}:
\begin{equation}
f_\text{eff} \frac{\langle \sigma v \rangle}{m_\chi} < 14 \frac{\text{pb}\cdot c}{ \text{TeV}},
\label{eq:IDcon}
\end{equation}
where $f_\text{eff}$ is an efficiency factor and $\langle \sigma v\rangle$ is the thermally-averaged DM annihilation rate. 
To zeroth order in DM velocity, the thermally-averaged cross-section for $\chi\chi\to Z_D Z_D$ is 
\begin{equation}
\langle{\sigma v}_0\rangle=\frac{4\pi \alpha_D^2}{m_\chi}\frac{(m_\chi^2-m_{Z_D}^2)^{3/2}}{(2m_\chi^2-m_{Z_D}^2)^2}.
\label{eq:}
\end{equation} 
Note that since the interaction responsible of DM annihilation involves only particles in the hidden sector, $\langle{\sigma v}_0 \rangle$ is independent of the kinetic mixing parameter $\epsilon$. 

As the DM particles are non-relativistic, Sommerfeld enhancements can become significant when $m_{Z_D}\ll m_\chi$ \cite{Sommerfeld:1931qaf, Hisano:2002fk, Hisano:2003ec, Hisano:2004ds}.
 We parameterize the effects as $\langle \sigma v \rangle \approx S_0 \langle \sigma v_0 \rangle$ \cite{Arkani-Hamed:2008hhe, Cirelli:2007xd, Tulin:2013teo}, following the treatment of \cite{Evans:2017kti}.
The Sommerfeld enhancement begins to be important for $m_\chi \sim 100$ GeV.

Since the leading process for DM annihilation is $s$-wave, we model the Sommerfeld enhancement to leading order in $v$ using the Hulth\'en potential \cite{Cassel:2009wt,Tulin:2013teo}:
\begin{equation}
S_0(\alpha,r,v)=\frac{2\pi \alpha_D}{v}\frac{\sinh{\left(\frac{6v}{\pi r}\right)}}{\cosh{\left(\frac{6v}{\pi r}\right)}-\cosh{\left(\sqrt{\left(\frac{6v}{\pi r}\right)^2-\frac{24\alpha_D}{r}}\right)}}
\label{eq:somer}
\end{equation}
where $r = m_{Z_D}/m_\chi$.

To evaluate the Sommerfeld enhancement at the epoch of recombination, we need the velocity of DM, $v$, during this epoch.
 To determine this velocity, we would need to compute the kinetic decoupling temperature of the DM, which depends on all four model parameters. 
 Here we take the simpler approach of providing a conservative estimate.

The first ingredient in this estimate is the HS kinetic decoupling temperature $\tilde T_\mathrm{kd}$. 
We approximate the HS kinetic decoupling temperature to be of order the HS freeze-out temperature, $\tilde T_\mathrm{fo}$: for HS mass ratios $m_{Z_D}\sim m_\chi/10$, the dark photon does not remain relativistic substantially after DM freeze-out, and so the freeze-out and the kinetic decoupling temperatures will be parametrically similar. 
Taking $\tilde T_\mathrm{fo}\approx \tilde T_\mathrm{kd}$ is a conservative approximation, since for hidden sectors with a larger hierarchy between dark photon and DM masses, where the Sommerfeld effect is most important, the dark photon bath will result in a lower value of $\tilde T_\mathrm{kd}$ and therefore a higher DM velocity at recoupling.
The second ingredient is the temperature ratio $\tilde T_\mathrm{fo}/T_\mathrm{fo}$, where $T_\mathrm{fo}$ is the SM freeze-out temperature. 
Numerically, we find that the coldest hidden sectors, i.e., model points with $\epsilon = \epsilon_\mathrm{min}$, are about 100 times colder than the SM at freeze-out, $\tilde T_\mathrm{fo}\sim 10^{-2} T_\mathrm{fo}$, varying with DM mass by a factor of order unity.
For the hottest hidden sectors, $\tilde T_\mathrm{fo} = T_\mathrm{fo}$. 

With $\tilde T_\mathrm{fo}\approx \tilde T_\mathrm{kd}$ and $v_\chi \sim 1$ at freeze-out, we then have 
\begin{equation}
v_\mathrm{CMB} \sim \frac{T_\mathrm{CMB}}{T_\mathrm{fo}} \sim \frac{T_\mathrm{CMB}}{m_\chi} \,\frac{10 \tilde T_\mathrm{fo}}{T_\mathrm{fo}}. 
\end{equation}
The first of the two factors varies from $10^{-10}$ to $10^{-12}$ for DM masses between 1 and 100 GeV, while the second factor varies within the range $0.1$ to 10 depending on the value of $\epsilon$. 
We find that the exclusion contours are not sensitive to the specific value of velocity adopted within this range. 
For definiteness we adopt the value $v_\mathrm{CMB}\approx 10^{-11}$ to show constraints.

The energy injection from DM annihilations is described by the redshift-independent total efficiency factor $f_\text{eff}$, which quantifies the efficiency for SM particles produced at a given energy to deposit their energy in the CMB plasma:
\begin{equation}
f_{\text{eff}}^{net}=\sum_l Br(Z_D \to \ell\ell)f_{\text{eff}}^{VV\to 4\ell}(m_\chi)+\sum_{X\neq l}Br(Z_D\to XX)f_{\text{eff}}^{XX}\left(\frac{m_\chi}{2}\right).
\label{eq:feff}
\end{equation}
We use the results from Ref.~\cite{Buschmann:2015awa} for the branching ratio of the dark photon and Ref.~\cite{Madhavacheril:2013cna} for the individual efficiency factors.
For non-leptonic channels, we evaluate $f_\text{eff}$ at $m_{\chi}/2$ to better account for the kinematic difference between direct and secluded annihilations. 
This is a good approximation since for hadronic channels the $f_\mathrm{eff}$ depend only weakly on $m_\chi$. 
For the $4\ell$ channels we make use of the dedicated calculation $f_\mathrm{eff}$ in \cite{Madhavacheril:2013cna}. 
These calculations are performed for DM masses as light as $m_\chi = 5$ GeV. 
To estimate the constraints on lighter dark matter masses, we extrapolate the result of the $4\ell$ channel down to $m_\chi = 1.5$ GeV which yields a value of $f_\text{eff} \sim 0.4$. 
We checked that the specific value of $f_\text{eff}$ does not affect our conclusions, and that a more involved calculation of this parameter should not change this number by more than a factor of 2.

%%%%%%%%%%%%%%%%%%%%%%%%
\begin{figure}[h!]
  \centering
  \includegraphics[width=0.7\linewidth]{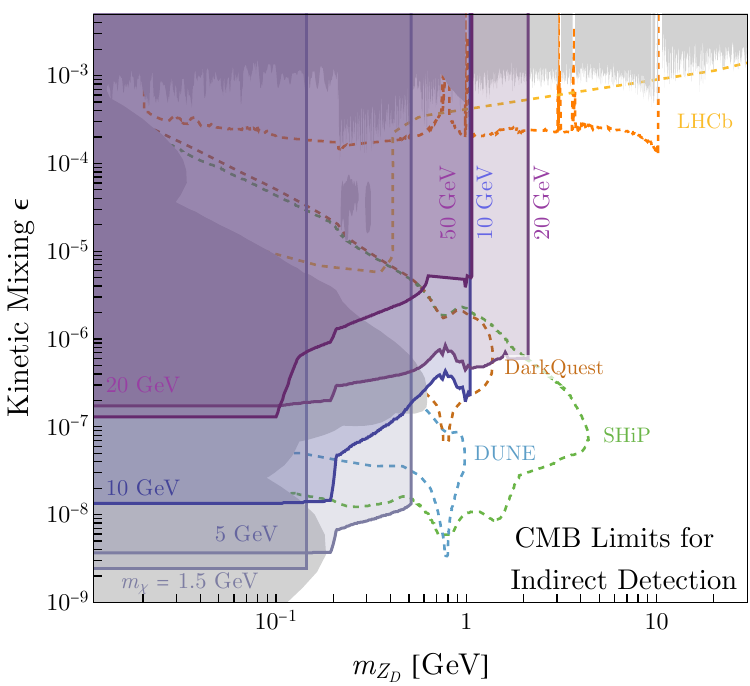}
  \caption{ Indirect detection constraints from CMB on the ($m_{Z_D}, \epsilon$) plane for several values of $m_\chi$ plotted with current and projected beam dump limits and other accelerator based experiments as LHCb and Belle II. 
  The CMB constraints are not monotonic in behavior as a function of the DM mass because of the Sommerfeld enhancements at larger masses. %are evaluated at a reference velocity $v= 10^{-11}$; see text for details.
  Additional curves showing excluded regions (shaded gray) and projected reach (colored dotted lines) from accelerator-based experiments are taken from Refs.~\cite{Batell:2022dpx, SHiP:2021nfo}, while exclusions from Supernova 1987a are taken from from Ref.~\cite{Chang:2016ntp}.  }
  \label{fig:indDet}
\end{figure}
%%%%%%%%%%%%%%%%%%%%%%%%

Imposing that the observed relic abundance of DM is obtained fixes $\alpha_D$ as a function of $m_\chi$ and $m_{Z_D}$.
Since we found this result to be approximately independent of the value $m_{Z_D}$ for dark photon masses sufficiently below $m_\chi$ (and $m_Z$), we take $\alpha_D$ to be a function only of the kinetic mixing parameter $\epsilon$ (for fixed $m_\chi$) and retain $m_{Z_D}$ dependence only as defined in equations Eqs.~\ref{eq:feff},~\ref{eq:somer}.
Only the out-of-equilibrium freeze-out and thermalized regions predict DM annihilation cross sections large enough to be constrained by indirect detection. 
We thus consider the portion of the relic curves with $\alpha_D (\epsilon) > \alpha_D(\epsilon_\mathrm{min})$, which determines a unique value of $\alpha_D$ at a given $\epsilon$.
The resulting bounds are shown in Fig.~\ref{fig:indDet} for several values of $m_\chi$.  
We see that, as expected, CMB constraints are strongest for the lightest DM masses, extending down to the smallest values of $\epsilon$ and thus to the smallest overall annihilation cross sections. 
For $m_\chi < 15$ GeV, the right boundary of the exclusion is set by the requirement that $m_{Z_D} < m_\chi/10$, as this is the regime where our calculations are self-consistent. 
As $m_\chi$ increases, the value of the excluded annihilation cross-section decreases, and thus the CMB constraints become successively weaker.
%Finally for $m_\chi = 100$ GeV the excluded parameter space increases again, as the Sommerfeld enhancement begins to kick in.
%
 Note that sizable portions of the projected sensitivity regions for DarkQuest, SHiP, and DUNE remain unconstrained\footnote{A similar conclusion holds for the FASER2 experiment \cite{Feng:2022inv}, which is not shown in the figure.}. 

%%%%%%%%%%%%%%%%%%%%%%%%%%%%%%%%%%%%%%%%%%%%%%%%%%%%%%%%%
%%%%%%%%%%%%%%%%%%%%%%%%
\subsection{Direct Detection}
\label{sec:DD}
%%%%%%%%%%%%%%%%%%%%%%%%

We now turn to direct detection constraints.
The cross-section for DM particles to scatter off of nuclei has contributions from both $Z$ and $Z_D$ exchange.
 The matrix element-squared for DM-nucleus scattering can be written as a function of recoil energy $E_R$, 
\begin{align}
\label{eq:erCS}
|\bar{\mathcal{M}}^{NR}(E_R)|^2&=A^2F^2(E_R)\left(\frac{f_{Z_D}}{m_{Z_D}^2+2m_N E_R}+\frac{f_Z}{m_Z^2}\right)^2, \\
\nonumber
 &\equiv A^2 F^2(E_R) |\hat{\mathcal{M}}(E_R)|^2
\end{align}
where $A$ is the mass number of the nucleus, $m_N$ is the mass of the nucleus, $F(E_R)$ is the Helm form factor \cite{Lewin:1995rx}, and $f_X$ encodes the various couplings to the partons, as shown in Tab. \ref{tab:couplings},
\begin{equation}
f_{(Z,Z_D)}=\frac{g_{(Z,Z_D)\chi }}{A}\left(Z(2g_{(Z,Z_D)u}+g_{(Z,Z_D)d})+(A-Z)(g_{(Z,Z_D)u}+2g_{(Z,Z_D)d})\right).
\end{equation}
In the second line of Eq.~\ref{eq:erCS}, we defined a reference per-nucleon matrix element-squared $|\hat{\mathcal{M}}(E_R)|^2$, as it will be useful in the following to separate the recoil energy dependence in the nuclear form factor from that in the partonic interaction.

The predicted event rate in direct detection experiments depends on the adopted DM halo profile as well as experiment-dependent recoil energy thresholds.
Following \cite{Evans:2017kti}, we define the observed event rate per unit detector mass in a given experiment as
\begin{equation}
R(\bar{\mathcal{M}}^{NR}(E_R)) = 
\frac{\rho_\chi}{2\pi m_\chi}\int_0^\infty \, dE_R \, |\bar{\mathcal{M}}^{NR}(E_R)|^2 \epsilon(E_R) \eta(E_R),
\end{equation}
where $\rho_\chi$ is the local halo density of DM particles near the Earth, which we take to be $\rho_\chi\sim 0.3$ GeV/cm$^3$, $\epsilon(E_R)$ is the efficiency of the detector as a function of recoil energy, and the mean inverse speed $\eta (E_R)$ can be written as \cite{Freese:2012xd}
\begin{align}
& \eta(E_R)=\int_{v_{\rm{min}}(E_R)}^{v_{\rm{esc}}}\frac{f(v)}{v}d^3v\simeq\frac{2\pi v_0^2}{N(v_0)}\exp{\left(-\frac{m_N E_R}{2\mu_{\chi N}^2v_0^2}\right)}, \\\nonumber
& N(v_0)=\pi^{3/2}v_0^3\left(\erf{\left(\frac{v_\mathrm{esc}}{v_0}\right)}-\frac{2}{\sqrt{\pi}}\frac{v_\mathrm{esc}}{v_0}\exp{\left(-\frac{v_\mathrm{esc}^2}{v_0^2}\right)}\right)
\end{align}
where the minimum velocity that can result in a recoil energy $E_R$ is given by $v_{\rm{min}}(E_R)=\sqrt{2m_NE_R}/2\mu_{\chi N}$ with $\mu_{\chi N}$ the DM-nucleus reduced mass. 
We have assumed the standard halo model, i.e., a Maxwell-Boltzmann distribution truncated at the galactic escape velocity, $f(v)=\exp{(-v^2/v_0^2)}/N(v_0)$, and, following \cite{Lin:2019uvt}, we take $v_\mathrm{esc}=550$ km/s and $v_0=220$ km/s. 
$N(v_0)$ is a normalization factor such that the function $f(v)$ integrates to one.

When $m_{Z_D}^2\gg 2m_N E_R$, the per-nucleon matrix element becomes effectively independent of the recoil energy. 
This enables us to immediately compare our calculated cross section with direct-detection experimental results, which are reported as constraints on a recoil-indepedent DM-nucleon cross section. 
For our model of interest, in the heavy-mediator regime, the DM-nucleon scattering cross section can be approximated as
\begin{equation}
\label{eq:sigma0}
\sigma^0_{\chi n} = \frac{\mu_{\chi n}^2|\hat{\mathcal{M}}(0)|^2}{\pi} =\frac{1}{\pi}\left(\frac{m_\chi m_n}{m_\chi+m_n}\right)^2\left(\frac{f_{Z_D}}{m_{Z_D}^2}+\frac{f_Z}{m_Z^2}\right)^2.
\end{equation}

For smaller values of $m_{Z_D}$, however, the DM-nucleon cross section will depend on the recoil energy. 
This complicates the comparison to published experimental constraints, since the shape of the predicted energy recoil spectrum in our model no longer matches the signal model considered by the experiment.
 In particular, for small values of $m_{Z_D}$, the signal spectrum peaks at lower recoil energies where it becomes more challenging to separate from various experimental backgrounds; see also \cite{Hambye:2018dpi}. %

We begin our assessment of direct detection constraints by quantifying the regions of the parameter space where the DD signal is well-described by a recoil-independent cross-section, and therefore where we can straightforwardly apply DD constraints on a $E_R$-independent spin-independent per-nucleon cross-section.
 To do so, we consider the ratio of observable event rates with and without partonic recoil energy-dependence:
\begin{equation}
\label{eq:ratio}
 \frac{R(\hat{\mathcal{M}}(E_R))}{R(\hat{\mathcal{M}}(0))}\equiv \frac{\sigma_{\chi n}^\mathrm{eff}}{\sigma^{0}_{\chi n}}.
\end{equation}
Here $\sigma_{\chi n }^\mathrm{eff}$ can be identified as an {\em effective} constant per-nucleon cross-section that would give rise to the same total event rate in a given experiment as the recoil-dependent model. 
It is worth emphasizing that this effective per-nucleon cross-section incorporates dependence on specific experimental efficiency thresholds, $\epsilon(E_R)$, as well as the choice of the halo model, $f(v)$. 
In the absence of background, the excluded value of this reference cross-section can be obtained straightforwardly by rescaling an experimental upper limit $\bar\sigma^{0}$ by the same ratio of observable event rates:
\begin{equation}
\label{eq:sigmaeff}
\bar \sigma_{\chi n}^\mathrm{eff}=\bar\sigma_{\chi n}^{0}\frac{R(\hat{\mathcal{M}}(E_R))}{R(\hat{\mathcal{M}}(0))} .
\end{equation}
In the presence of backgrounds, however, constraints depend on the shape of the recoil-energy spectrum, not just the total event rate. 
In our model, points with substantial recoil energy dependence have a spectrum that is more challenging to separate from backgrounds, and thus in general experimental exclusions on the recoil-dependent model will be weaker than the exclusion given by the simple rescaling of Eq.~\ref{eq:sigmaeff}.

The ratio $\sigma_{\chi n}^\mathrm{eff}/\sigma^{0}_{\chi n}$ in Eq. \ref{eq:ratio} usefully encapsulates the impact of the recoil energy dependence on the signal spectrum at a given experiment.
%
%%%%%%%%%%%%%%%%%%%%%%%%%
\begin{figure}[t!]
 \centering
 \includegraphics[width=0.7\linewidth]{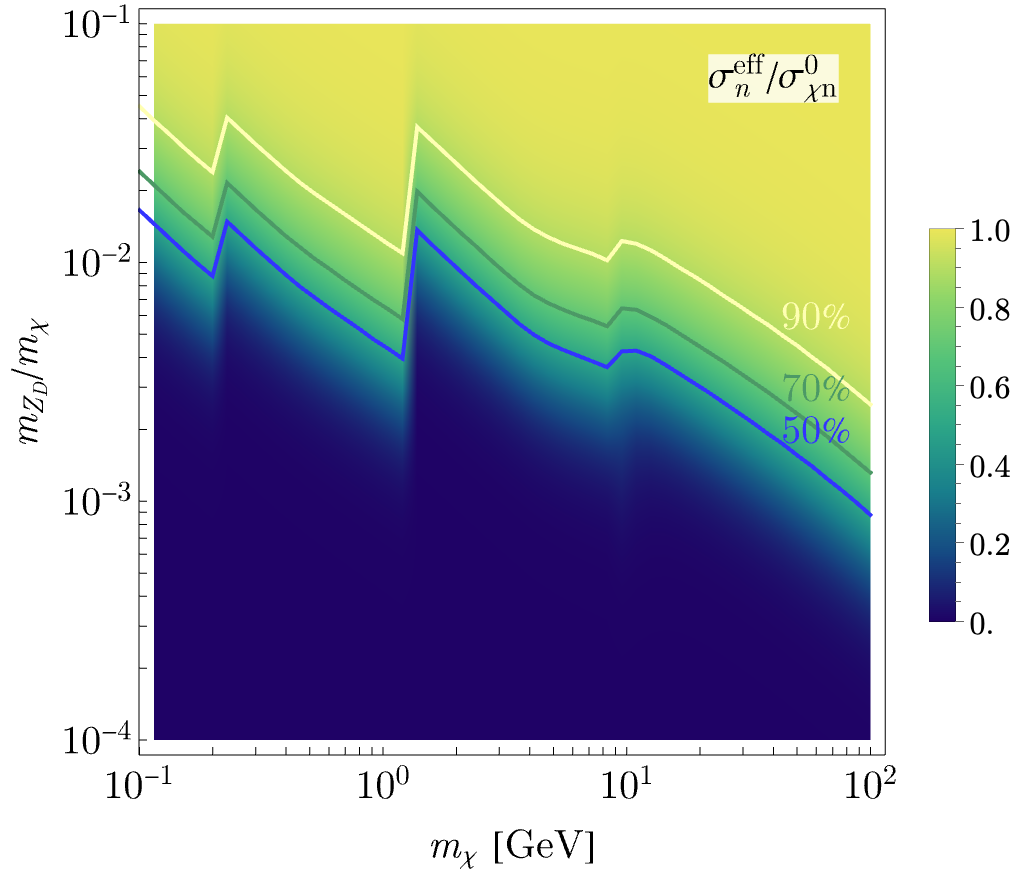} 
 \caption{The ratio $\sigma_{\chi n}^\mathrm{eff}/\sigma^{0}_{\chi n}$ as a function of $m_\chi$ and $m_{Z_D}/m_\chi$, showing the transition from regions where the DM-nucleon scattering can be reasonably approximated as constant (yellow) to regions where the recoil energy dependence is sizable (blue).
We show contours at 90\%, 70\%, and 50\%. 
 Jumps at various values of $m_\chi$ reflect transitions between different experiments as discussed in the text.}
 \label{fig:ratio}
\end{figure}
%%%%%%%%%%%%%%%%%%%%%%%%%%
%
We show the value of $\sigma_{\chi n}^\mathrm{eff}/\sigma^{0}_{\chi n}$ in Fig.~\ref{fig:ratio} as a function of $m_\chi$ and $m_{Z_D}/m_\chi$. 
Here we have used the recoil energy efficiencies $\epsilon (E_R)$ from the experiments that currently provide the most stringent constraints on $\sigma^{0}_{\chi n}$ in a given DM mass range \cite{CRESST:2024cpr, CRESST:2019jnq,DarkSide-50:2022qzh,XENON:2024hup,LZCollaboration:2024lux}. 
The jumps seen in $\sigma_{\chi n}^\mathrm{eff}/\sigma^{0}_{\chi n}$ occur as we move from one experiment to the next. In particular, we use results from the following experiments in the indicated mass ranges: CRESST 2024 in ($0.1\sim 0.2$) GeV \cite{CRESST:2024cpr}; CRESST 2019 in ($0.2\sim 1.2$) GeV \cite{CRESST:2019jnq}; DarkSide-50 in ($1.2\sim 4$) GeV \cite{DarkSide-50:2022qzh}; XENONnT in ($4\sim 9$) GeV \cite{XENON:2024hup}; and LZ in ($9\sim 100$) GeV \cite{LZCollaboration:2024lux}. 
For each DM mass, the heaviest dark photons we consider lead to a constant per-nucleon cross-section, while the lightest ones do not. 
We show three contours indicating where the effective per-nucleon cross-section is 90\%, 70\%, and 50\% of the zero-recoil value. 

%%%%%%%%%%%%%%%%%%%%%%%%%%
\subsubsection{Direct-Detection Constraints to Beam-Dump Parameter Space} 
%%%%%%%%%%%%%%%%%%%%%%%%%%

%%%%%%%%%%%%%%%%%%%%%%%%%
\begin{figure}[t!]
 \centering
 \includegraphics[width=0.38\linewidth]{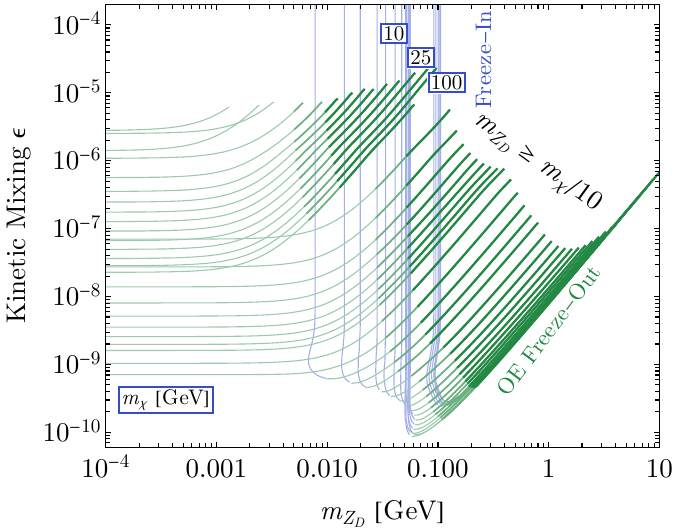} \ \ \ \ \
 \includegraphics[width=0.38\linewidth]{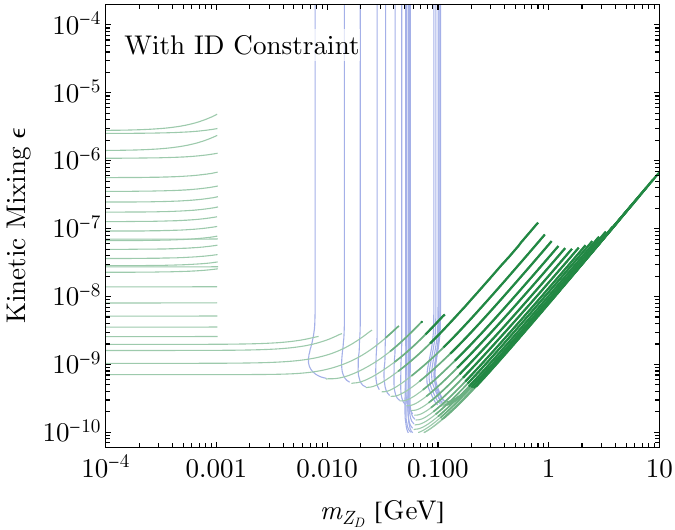}
 \includegraphics[width=0.38\linewidth]{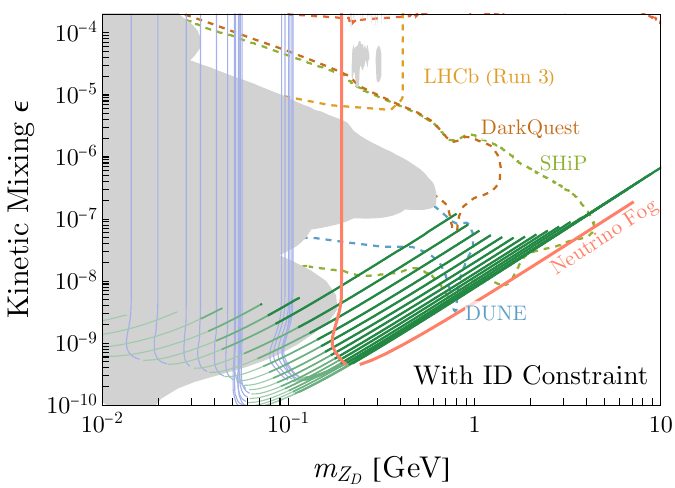}\ \ \ \ \
 \includegraphics[width=0.38\linewidth]{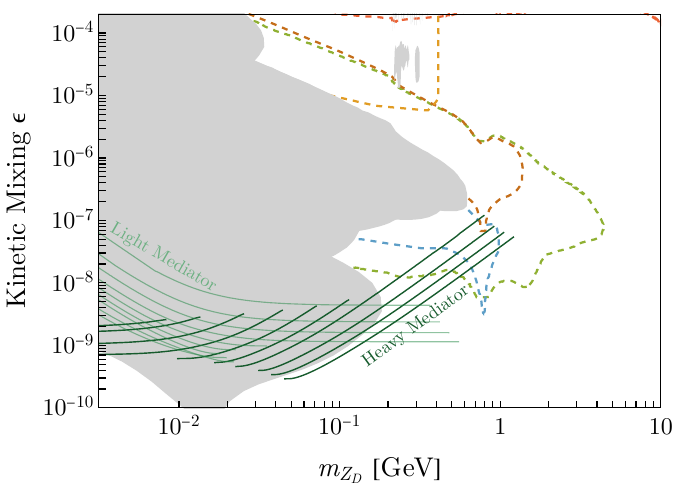}
 \caption{The successive visualization of direct detection (DD) constraints, combined with indirect detection (ID) exclusions, in the dark photon parameter space.
 Top left: contours corresponding to the maximal allowed DD cross-section are drawn for fixed DM masses using the relationship between $\alpha_D$ and $\epsilon$ determined by the relic abundance curves, with colors corresponding to different regimes as indicated. Contours are labeled with DM mass in GeV. We require $m_{Z_D}< 0.1m_\chi$. 
 Regions above and to the left of each contour are excluded for the corresponding DM mass. Thick, medium, and thin line weights indicate $\sigma_{\chi n}^\mathrm{eff}/\sigma^{0}_{\chi n}>0.9$, 0.7, and 0.5, respectively, as discussed in the text.
 Top right: we impose CMB constraints set by Eq.~\ref{eq:IDcon}. 
 Bottom left: we add the current and projected constraints from SN1987a and beam-dump experiments, as well as projections for LHCb. 
 Additionally the orange line shows the neutrino fog lines for a 100 GeV DM particle.
Bottom right: We compare the DD contours shown in the top left panel for DM masses in the range $3\,\gev <m_\chi < 12\,\gev$ with several contours obtained using XENONnT results for DM of the same masses scattering through light mediators \cite{XENON:2024hup}. 
}
\label{fig:DDToBeam_0}
\end{figure}
%%%%%%%%%%%%%%%%%%%%%%%%%%

As discussed in Sec. \ref{subsec:prelim}, our model is defined by four parameters: $m_\chi$, $m_{Z_D}$, $\epsilon$, and $\alpha_D$, while experimental searches for dark photons are most naturally characterized by the two independent parameters $(m_{Z_D}, \epsilon)$.
 For each value of $m_\chi$, we can use the relic abundance constraint to determine $\alpha_D(\epsilon)$.
  Again, since $\alpha_D$ is not a monotonic function of $\epsilon$, it is useful to separately discuss the freeze-in region $\alpha_D < \alpha_{D}(\epsilon_\mathrm{min})$ and the freeze-out regions $\alpha_D > \alpha_{D}(\epsilon_\mathrm{min}$). 
  Given $m_\chi$ and $\alpha_D(\epsilon)$, the corresponding direct-detection cross section can then be obtained for a given value of $(m_{Z_D}, \epsilon)$, for values of those parameters that are physically consistent within the model. 

We compare the effective DD cross section, $\sigma_{\chi n}^{\rm{eff}}$, to the observed limit $\bar\sigma^0$ on the per-nucleon cross-section for each value of $m_\chi$. 
Setting $\sigma^\mathrm{eff}_{\chi n} = \bar\sigma^0$ then identifies the values of $(m_{Z_D}, \epsilon)$ that give a DD cross section that saturates the upper limit.
This procedure results in an accurate constraint on the parameter space of the model when the partonic recoil energy dependence is small (see yellow region in Fig.~\ref{fig:ratio}), but overestimates the exclusions when the recoil energy dependence is sizable (blue region in the figure).

The bound is shown in the top left panel of Fig.~\ref{fig:DDToBeam_0} for DM masses in the range $(0.5,100)$ GeV. 
The line weights of the curves are thick for regions with $\sigma_{\chi n}^\mathrm{eff}/\sigma^{0}_{\chi n}>0.9$, medium for $\sigma_{\chi n}^\mathrm{eff}/\sigma^{0}_{\chi n}>0.7$, and thin for all other regions. 
Since the DD cross section increases with increasing $\epsilon$ and decreasing $m_{Z_D}$, regions above and to the left of the curve are excluded for the corresponding DM mass. 
Here the blue portion of the curve indicates the freeze-in regime, while the green portion of the curve denotes the out-of-equilibrium freeze-out regime. 
These two different regimes predict different values of $\alpha_D(\epsilon)$, and thus must be independently constrained, i.e., the freeze-in regime is not excluded by the out-of-equilibrium freeze-out constraint.
In the top right panel of Fig.~\ref{fig:DDToBeam_0} we additionally impose the CMB constraints derived in the previous subsection. 
These constraints eliminate a large fraction of the curves in the out-of-equilibrium freeze-out regime for lower dark matter masses.
In the bottom left panel, we superimpose the constraints on the dark photon parameter space from beam dump and supernova bounds (gray regions) and indicate the projected sensitivity of future (beam dump and LHCb) experiments. 
We can conclude that visibly-decaying dark photons, consistent with current DD constraints, are discoverable at beam dump experiments.
Note that in the regions relevant for future beam dump experiments, the direct detection exclusions in the out-of-equilibrium freeze-out regime are well characterized by the constant cross-section approximation. 
Also in this bottom left panel, we show a visualization of the neutrino fog in beam dump parameter space, indicated in orange. 
We use the neutrino fog cross-section for a 100 GeV DM particle scattering off of xenon, with the LZ recoil energy thresholds, taken from \cite{OHare:2021utq}, and map it into dark photon parameter space following the same procedure as for the exclusions.
We show only portions of this line that satisfy $\sigma_{\chi n}^\mathrm{eff}/\sigma^{0}_{\chi n}>0.7$, and in fact almost all of the out-of-equilibrium portion of the line satisfies $\sigma_{\chi n}^\mathrm{eff}/\sigma^{0}_{\chi n}>0.9$. 
Regions of parameter space below and to the right of the orange line lie below the neutrino fog, hence are challenging to access in direct detection experiments. 
We show the fog for a 100 GeV DM particle: this is the heaviest DM mass we consider, and it accordingly yields the maximal amount of parameter space above the neutrino fog in most of parameter space (although for lighter DM masses, regions of interest above the fog can extend to smaller values of $m_{Z_D}$ and $\epsilon$ in the out-of-equilibrium freeze-out regime). 
As evident in the figure, there is a notable chunk of parameter space in the out-of-equilibrium freeze-out regime that is inaccessible to intensity frontier experiments, but yields direct detection signals above the neutrino fog.

Finally in the bottom right panel, we compare the DD bounds we obtain from this rescaling procedure (here shown with a heavy line weight for all values of $\sigma_{\chi n}^\mathrm{eff}/\sigma^{0}_{\chi n}$) to a set of bounds obtained using the light-mediator results from XENONnT \cite{XENON:2024hup}, shown with a light line weight.
We show results for the subset of DM masses $3\,\gev <m_\chi < 12\,\gev$ where XENONnT's light mediator results apply. 
Broadly, we expect the rescaling procedure to work well when $m^2_{Z_D} \gtrsim 2 m_N E_R$, while the XENONnT light mediator results apply when $m^2_{Z_D}\lesssim 2 m_N E_R$. 
Thus we expect the full DD upper bound on a given DM mass to follow the corresponding heavy and light contours at the largest or smallest values of the dark photon mass, respectively, and to interpolate between them in the area where these two contours cross. 
The primary conclusion to draw from this comparison is that the correct DD bounds in the light mediator regime lie deeply within excluded regions of dark photon parameter space for DM masses $m_\chi < 12\,\gev$. 
For heavier DM masses, we expect the turnover region, where $m_{Z_D}\sim 2 m_N E_R $, may lie in open regions of dark photon parameter space, but the light-mediator regime will still be deeply excluded by constraints on dark photons. 
Since the turnover region is substantially below the projected sensitivities of next-generation beam dump experiments, it will not be useful here to treat the recoil-energy-dominated region in more detail.

Of the projected beam dump experiments shown in Fig.~\ref{fig:DDToBeam_0} (see bottom left panel), DarkQuest and SHiP are sensitive to the largest values of $\epsilon$. 
In this portion of parameter space, visibly-decaying dark photons are consistent with this minimal dark matter model mainly in the freeze-in regime.
SHiP and DUNE are also sensitive to more weakly-coupled dark photons, and thus visibly-decaying dark photons discoverable in SHiP and DUNE can be consistent with out-of-equilibrium freeze-out scenarios as well as freeze-in. 
The dark photons discoverable by LHCb, meanwhile, are only consistent with direct and indirect detection constraints in the freeze-in regime. Successful freeze-in for these dark photons requires very small dark gauge couplings, $\alpha_D \lesssim 10^{-17}$.

%%%%%%%%%%%
\subsubsection{Beam-Dump Reach in Direct-Detection Parameter Space} 
%%%%%%%%%%%

We now turn to mapping beam dump sensitivities into the DD parameter space, which is most naturally characterized by the set of parameters $(m_\chi, \sigma_{\chi n}^\mathrm{eff})$. 
The effective DD cross-section $\sigma_{\chi n}^\mathrm{eff}$ depends on all four model parameters. 
The mapping that we describe in this section is not a one-to-one mapping, since there exist multiple pairs of $(\epsilon, m_{Z_D}$) that lead to the same direct detection cross section.

We can visualize the reach of searches for visibly-decaying dark photons in the DD plane via a two-step procedure. 
First, we use the relic abundance constraint to determine $\alpha_D(\epsilon)$ for each value of $m_\chi$. 
We find it useful to separately analyze the freeze-in region $\alpha_D(\epsilon) < \alpha_{D}(\epsilon_\mathrm{min})$, the out-of-equilibrium freeze-out region with $\alpha_{D}(\epsilon_\mathrm{min})<\alpha_D(\epsilon) <\alpha_D^{TF} $, as well as the region above the thermalization floor, $\alpha_D(\epsilon) =\alpha_D^{TF}$.

Second, for fixed $m_\chi$, we use each value of $\alpha_D$ to determine all values of $\sigma_{\chi n}^\mathrm{eff}$ that can be obtained given the $(m_{Z_D}, \epsilon)$ within each beam dump experiment's and LHCb's projected reach. We only consider model points that are not already excluded by either direct dark photon searches or CMB constraints (see Fig.~\ref{fig:indDet}).
We restrict our treatment to model points that satisfy $\sigma_{\chi n}^\mathrm{eff}/\sigma^{0}_{\chi n}>0.7$, to more accurately characterize the sensitivity of DD experiments. 
We also do not include dark photon masses that are within 75 MeV of the $\rho$-meson mass, where resonant $Z_D$-$\rho$ mixing allows beam dump experiments to extend their sensitivity to much smaller values of $\epsilon$. 
This choice is motivated by the fact that the Boltzmann equations we discussed in Sec. \ref{sec:boltz} do not take into account the $Z_D$-$\rho$ mixing. 
In this region, additional care is needed to model $Z_D$-$\rho$ mixing even at zero temperature \cite{Kamada:2024ntk,LoChiatto:2024guj}. 
An accurate treatment of the DM relic abundance calculation with $m_{Z_D}\approx m_\rho$ also requires a dedicated consideration of resonantly enhanced mixing at finite temperature, which we leave to future work.

\begin{figure}[t!]
  \centering
  \includegraphics[width=0.36\linewidth]{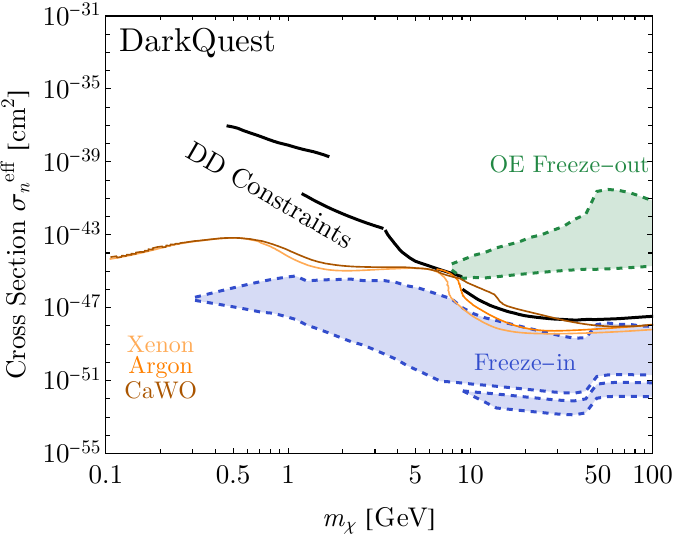} 
  \includegraphics[width=0.36\linewidth]{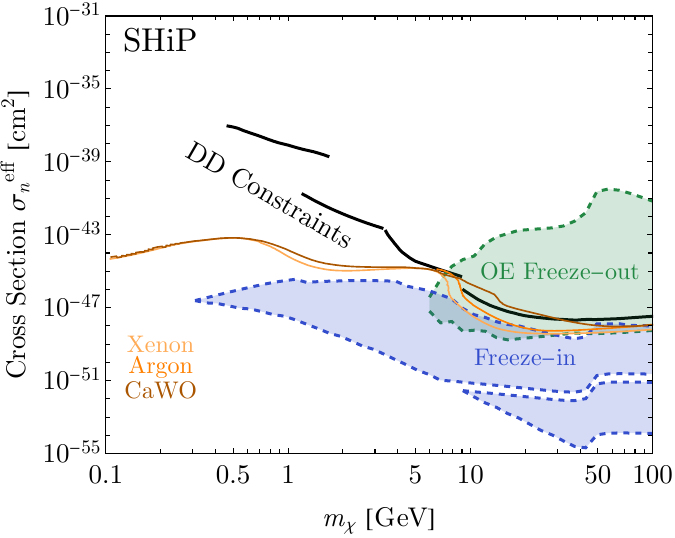} 
  \includegraphics[width=0.36\linewidth]{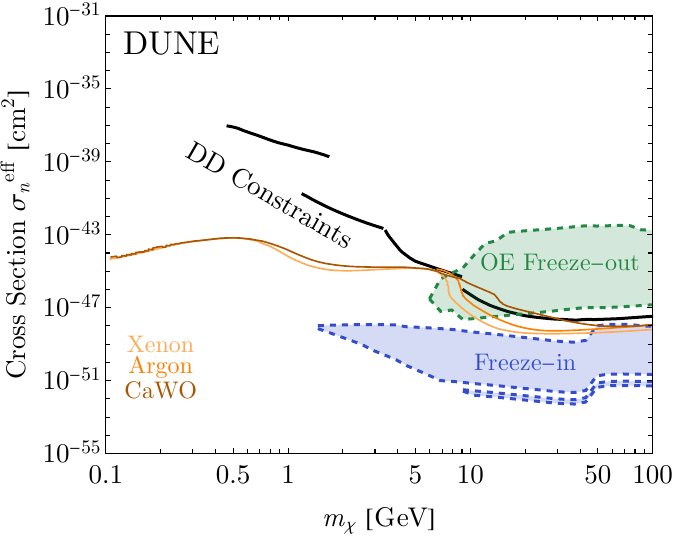} 
  \includegraphics[width=0.36\linewidth]{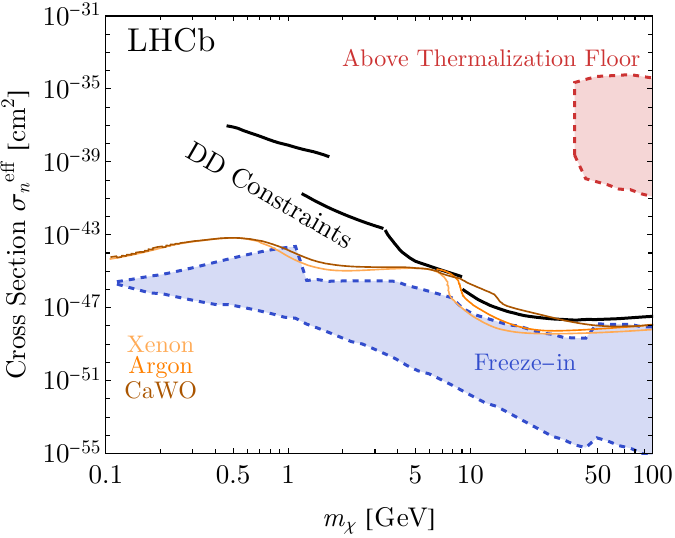} 
  \caption{Regions of parameter space that will be probed by DarkQuest, SHiP, and DUNE in the DD plane $(m_\chi, \sigma_{\chi n}^{\rm{eff}})$.
The three colors correspond to regions that are above the thermalization floor (in red), out-of-equilibrium freeze-out (green), and freeze-in (blue). 
Only parameter points consistent with CMB exclusions are shown, while current DD constraints are shown as black lines. 
The orange lines indicate the neutrino fog for several materials \cite{OHare:2021utq}.}
 \label{fig:Beam_to_DD}
\end{figure}

The procedure outlined above leads to three regions in the $(m_\chi, \sigma_{\chi n}^{\mathrm{eff}})$ plane.
We color these regions to distinguish three regimes: red regions are above the thermalization floor, the green ones are in the out-of-equilibrium freeze-out regime, and blue regions correspond to freeze-in. 
The results are shown in Fig.~\ref{fig:Beam_to_DD}. We also show the neutrino fog for several materials from Ref.~\cite{OHare:2021utq}, including that for CaWO$_4$, which we expect to be roughly representative of the neutrino fog for Al$_2$O$_3$ in the DM mass range where scattering on oxygen dominates the DM signal. 

As can be seen from Fig.~\ref{fig:TFFIcontours}, the sensitivity of any given beam dump experiment and LHCb extends to some minimum value of $\epsilon$. 
This minimum value of $\epsilon$ is substantially above the value that separates the freeze-in and freeze-out regimes.
 Thus at a specific value of $m_\chi$, there is a gap between the values of $\alpha_D$ attained in the freeze-in region and those attained in the freeze-out regimes. 
 This gap accounts for the separations of the freeze-in regions from the other regions in Fig.~\ref{fig:Beam_to_DD} (most notable for LHCb). 
 The gap visible within each freeze-in region comes from the $\rho$-window cut.\footnote{This gap is not visible in the out-of-equilibrium freeze-out region, since in this region the DD cross section and the DM relic abundance depend on different combinations of parameters.}

%%%%%%%%%%%%%%%%%%%%%%%%%%%%%%%
%%%%%%%%%%%%%%%%%%%%%%%%%%%%%%%

Several comments for the three cosmological regimes are in order: 
\begin{itemize}
\item The combination of direct and indirect detection constraints rule out the above-the-thermalization floor scenario for all experiments. 
As already seen in Fig.~\ref{fig:TFFIcontours}, the parameter space probed by DUNE, DarkQuest, and SHiP lies below the thermalization floor for the considered masses. Here we also see that in the parameter space probed by LHCb, direct detection constraints leave freeze-in as the only viable regime.

\item At low DM masses, SHiP and DUNE can probe portions of the out-of-equilibrium regime that are not probed by DD. 
This is not the case for DarkQuest, once we exclude the regions where the dark photon mass is within 75 MeV of the $\rho$ mass, nor the case for LHCb.

\item Beam dumps can probe parameter space consistent with direct detection cross sections that are well below the neutrino fog for both the out-of-equilibrium and the freeze-in regimes. The freeze-in regime below the neutrino fog can be probed by LHCb, as well. 

\item All three beam dump experiments and LHCb can probe regions of the freeze-in regime that are well beyond the exclusions of DD experiments.
\end{itemize}
It is worth emphasizing that the colored regions in Fig.~\ref{fig:Beam_to_DD} refer only to the portion of dark photon parameter space that is covered by specific (proposed) experiments, which is only a subspace of the viable parameter space of the model.

%%%%%%%%%%%%%%%%%%%%%%%%%%%
\section{Discussion and Conclusions}
\label{sec:conclusions}
%%%%%%%%%%%%%%%%%%%%%%%%%%%
%
In this paper, we analyzed a secluded DM model where the DM is a Dirac fermion that interacts with the SM through the exchange of a dark photon.  We focus on the DM mass range GeV $\lesssim m_\chi \lesssim 100$ GeV: light enough that DM production in the early universe is dominated by temperatures below electroweak symmetry breaking, and heavy enough for nuclear recoil, rather than electron recoil \cite{DAMIC-M:2025luv}, to dominate direct detection constraints. Within the considered area of parameter space, we find that regions above the thermalization floor (the ``WIMP-next-door'' regime) are now entirely excluded through a combination of CMB and direct detection constraints. 
However, there is still interesting parameter space for this model {\em below} the thermalization floor. 
This parameter space can be broadly split into two regimes: one where the thermal history of the model is dominated by an interplay of DM annihilations to dark photons with the out-of-equilibrium production of DM pairs from the SM, which we refer to here as ``out-of-equilibrium freeze-out'', and one where the DM relic abundance is fixed through freeze-in from the SM. 

One important result of this work is that a discovery of a visibly-decaying dark photon at future beam-dump experiments is consistent with this very minimal model of dark matter.
 In particular, this model still has viable parameter space for the proposed experiments DUNE, SHiP, and DarkQuest.
 In a large portion of this parameter space the DM attains its relic abundance through the relatively simple freeze-in process, but there are also surviving regions where the more involved out-of-equilibrium freeze-out scenario controls the DM abundance. 
A discovery could be made at LHCb, but only for dark matter in the freeze-in region. If a dark photon were discovered at a collider experiment, the model would also predict a small but non-zero dark matter pair production cross section, which given the dark photon parameters would be determined up to the dark matter mass. This signature would be a challenging but definite target for the LHC and future colliders, which would provide complementary information on the connection between the observed dark photon and dark matter. Similarly, a dark photon discovery would provide motivation to pursue DD sensitivity below the neutrino fog.

Meanwhile for nuclear-recoil direct detection experiments, the remaining discovery sensitivity to this specific model lies primarily in the out-of-equilibrium freeze-out regime. 
In particular, there remains open parameter space in this regime, with DD cross-sections above the neutrino fog, that is inaccessible to planned beam-dump experiments. 
In this part of parameter space, future DD results offer the best discovery prospects for this model. 
Where the dark photons are heavy enough that their mass dominates over the recoil energy in DM-nucleon scattering, the bulk of the open model parameter space does lie below the neutrino fog, however.
Meanwhile for lighter dark photons, we have used XenonNT results for light mediators~\cite{XENON:2024hup} to show that, within the context of this minimal model, such light-mediator DD searches are generically sensitive to portions of dark photon parameter space that are already ruled out. 

It is worth bearing in mind that all the terrestrial and astrophysical constraints on visibly-decaying dark photons that we have quoted are subject to  uncertainties.  At beam dumps, updated calculations of and systematic uncertainties in various dark photon production processes can notably impact experimental sensitivities \cite{Curtin:2023bcf,Kyselov:2024dmi,Zhou:2024aeu}.
 Supernova constraints, meanwhile, are sensitive to uncertainties in the detailed modeling of the progenitor, in particular its mass. Additionally, constraints on energy deposition in the stellar envelope from BSM particles can strengthen the supernova constraints on the production of light feebly-interacting particles \cite{Falk:1978kf,Sung:2019xie,Shin:2022ulh,Caputo:2022mah}. 
 The resulting constraints from SN1987a computed in~\cite{Sung:2019xie} for dark photons are accordingly somewhat more stringent than what we show here given their fiducial modeling of the progenitor mass. In particular these bounds extend the excluded area of the dark photon parameter space in the region where the dark photon is appreciably light in direct detection experiments, i.e., for $\sigma_{\chi n}^\mathrm{eff}/\sigma^{0}_{\chi n}\lesssim 0.5$.
 However our main conclusions about the complimentary discovery sensitivity of both beam dump and DD experiments are not affected by the uncertainties in current constraints on visibly-decaying dark photons.

Since potential connections to dark matter provide some of the best theoretical motivations for MeV-GeV scale dark photons, it is important to understand the potential cosmological consequences of discovering a visibly-decaying dark photon. 
Here we have shown that dark photon discovery at the next generation of intensity frontier experiments is consistent with a very minimal model of thermal dark matter. 
It is worth commenting that this minimal model admits a range of extensions that further expand its parameter space. For example, the model could contain a dark Higgs boson.
In this paper, we invoked a St\"uckelberg origin for the dark photon mass, but
the phenomenology of the model in the regime of interest does not in fact depend strongly on this choice, and is also compatible with a dark Higgs mechanism.
In particular, the DM in this model already has an $s$-wave annihilation cross-section and a leading spin- and velocity-independent contribution to the DM-nucleon cross-section. 
Thus the predictions of the minimal model are largely unchanged by the addition of a physical dark Higgs boson in the spectrum, provided that the vector portal coupling to the SM dominates the energy injection into the hidden sector compared to any Higgs-portal coupling resulting from dark Higgs-SM Higgs mixing.
 Cosmologically, requiring the vector portal to dominate over the Higgs portal can lead to problematically long dark Higgs boson lifetimes if the dark Higgs has only SM decay modes available. The resulting disruption to BBN can readily be avoided by ensuring that the dark Higgs is heavy enough to decay to pairs of dark photons. 

More consequentially, adding a small $U(1)_D$-breaking contribution to the DM mass splits the Dirac DM into two Majorana states, of which only the lightest is absolutely stable (see, e.g., \cite{Tucker-Smith:2001myb,Alexander:2016aln, CarrilloGonzalez:2021lxm}).
This inelastic deformation can parametrically suppress  direct detection cross-sections, which will generally serve to enhance the relative discovery potential of visible dark photon searches, while introducing additional cosmological signatures in the freeze-in regime \cite{Heeba:2023bik}. 

Ultimately, we come to the important conclusion that this minimal model of hidden sector DM remains viable and  within reach of near-future terrestrial experiments.  The powerful constraints from current direct and indirect experiments push this hidden sector DM model into regimes where they are out of equilbrium with the SM in the early universe.  Out of equilibrium, the predictions of even a minimal DM model can be involved.  Here we have demonstrated, within this simple and minimal model, how the various established methods of searching for feebly-interacting particles (direct and indirect DM detection as well as beam-dump searches for dark mediators) provide complementary discovery sensitivity in the open parameter space of the model. 

\medskip

\paragraph{Acknowledgements.}
We thank Julian Mu\~noz, Tracy Slatyer, Katelin Schutz, Jesse Thaler, and Douglas Tuckler for useful conversations. 
The work of AA and JS was supported in part by U.S. Department of Energy (DOE) Office of High Energy Physics under grant numbers DE-SC0023365 and DE-SC0015655. The work of AA was supported in part by Kuwait University. The work of CC is supported in part by DOE grant DE-SC0012567 and in part by the Simons Foundation (Grant Number 929255, T.R.S).
 The research of SG is supported in part by DOE grant number DE-SC0010107. JS additionally thanks the Massachusetts Institute of Technology as well as the Kavli Institute for Theoretical Physics (KITP), supported in part by grant NSF PHY-2309135, for hospitality during the performance of this work.

%%%%%%%%%%%%%%%%%%%%%%%%%%%%%%%%%%%%%%%%%%%%%%%%%%%%%%%%%

\appendix

\section{Cross Section Expressions}
\label{app:CrossSec}
In this section we give the explicit form for the functions $\mathcal{A}(s)$, defined in Eq.~\ref{eq:As} as the amplitude-squared for $2\to 2$ scatterings, summed over final state and averaged over initial state spins and integrated over final state phase space, for the processes $f\bar f\to \chi\bar \chi$ and $ \chi\bar \chi\to Z_D Z_D$.  

 \paragraph{SM fermion annihilation.}
The integrated spin averaged amplitude-squared for this process is
\begin{equation}
\begin{aligned}
\mathcal{A}_{f\bar f\to \chi\bar \chi}(s) = \frac{1}{g_{f}g_{\bar{f}}}\frac{8\alpha_D}{3} & \sqrt{1-\frac{4m_\chi^2}{s}} \ (2m_\chi^2+s)\Bigg( \left(\frac{C_A^2(s - 4m_f^2) + C_V^2 (2m_f^2 + s)}{(s - m_Z^2)^2 +m_Z^2 \Gamma_Z^2 }\right) \\
+ & \left(\frac{\left({C'}_A C_A(s - 4m_f^2) + {C'}_V C_V (2m_f^2 + s)\right)(s - m_Z^2)}{((s - m_Z^2)^2 +m_Z^2 \Gamma_Z^2)(s-m_{Z_D}^2) }\right) \\
+ &\left(\frac{{C'}_A^2(s - 4m_f^2) +{C'}_V^2 (2m_f^2 + s)}{(s - m_{Z_D}^2)^2}\right)
\Bigg)
\end{aligned}
\end{equation}
where $C^{(')}_A$, $C^{(')}_V$ are the axial and vectorial couplings of the SM fermions to the $Z$ ($Z_D$) respectively. 
These couplings can be obtained from Tab. \ref{tab:couplings} and are given by
\begin{equation}
\begin{aligned}
C_A = & -\frac{1}{2} \frac{g}{\cos\theta_W}\left( \cos\alpha \cos^2 \theta_W T^3 + (\eta \sin\alpha - \cos\alpha \sin\theta_W)(Y_L-Y_R)\sin\theta_W \right)\\
C_V = & \frac{1}{2} \frac{g}{\cos\theta_W}\left( \cos\alpha \cos^2 \theta_W T^3 + (\eta \sin\alpha - \cos\alpha \sin\theta_W)(Y_L+Y_R)\sin\theta_W \right)\\
C'_A = & \frac{1}{2} \frac{g}{\cos\theta_W}\left( \sin\alpha \cos^2 \theta_W T^3 - (\eta \cos\alpha + \sin\alpha \sin\theta_W)(Y_L-Y_R)\sin\theta_W \right) \\
C'_V = &- \frac{1}{2} \frac{g}{\cos\theta_W}\left( \sin\alpha \cos^2 \theta_W T^3 - (\eta \cos\alpha + \sin\alpha \sin\theta_W)(Y_L+Y_R)\sin\theta_W \right)
\end{aligned}
\end{equation}
where $g$ is the weak coupling, and $T^3, Y_L, Y_R$ are the third component of the isospin and the hypercharge of the left handed and right handed fermion, respectively. 
Note that there are no axial charges for the hidden sector fermions as, by construction, they have equal charges for the left- and right-handed fields. 

\paragraph{DM annihilation to dark photons.}
The integrated amplitude-squared for this process is~\cite{Fernandez:2021iti},
\begin{equation}
\mathcal{A}_{\chi\bar\chi\rightarrow Z_D Z_D}(s) = 8\pi \alpha_D^2\sqrt{1-\frac{4m_\chi^2}{s}} \left( 2\frac{(s^2+4 s m_\chi^2 - 8 m_\chi^4)}{s (s-4m_\chi^2)}\tanh^{-1}\left(\sqrt{\frac{s-4m_\chi^2}{s}}\right)-\frac{(s+4m_\chi^2)}{\sqrt{s(s-4m_\chi^2)}}\right).
\end{equation}

%%%%%%%%%%%%%%%%%%%%%%%%%%%%%%%%%%%%%%%%%%%%%%%%%%%%%%%%%
\bibliographystyle{JHEP}
\bibliography{floor}

\providecommand{\href}[2]{#2}\begingroup\raggedright\begin{thebibliography}{10}

\bibitem{Abercrombie:2015wmb}
D.~Abercrombie et~al., {\it {Dark Matter benchmark models for early LHC Run-2
  Searches: Report of the ATLAS/CMS Dark Matter Forum}},  {\em Phys. Dark
  Univ.} {\bf 27} (2020) 100371, [\href{http://xxx.lanl.gov/abs/1507.0096}{{\tt
  arXiv:1507.0096}}].

\bibitem{Alexander:2016aln}
J.~Alexander et~al., {\it {Dark Sectors 2016 Workshop: Community Report}},  8,
  2016.
\newblock \href{http://xxx.lanl.gov/abs/1608.0863}{{\tt arXiv:1608.0863}}.

\bibitem{Battaglieri:2017aum}
M.~Battaglieri et~al., {\it {US Cosmic Visions: New Ideas in Dark Matter 2017:
  Community Report}},  in {\em {U.S. Cosmic Visions: New Ideas in Dark
  Matter}}, 7, 2017.
\newblock \href{http://xxx.lanl.gov/abs/1707.0459}{{\tt arXiv:1707.0459}}.

\bibitem{Gori:2022vri}
S.~Gori et~al., {\it {Dark Sector Physics at High-Intensity Experiments}},
  \href{http://xxx.lanl.gov/abs/2209.0467}{{\tt arXiv:2209.0467}}.

\bibitem{Antel:2023hkf}
C.~Antel et~al., {\it {Feebly-interacting particles: FIPs 2022 Workshop
  Report}},  {\em Eur. Phys. J. C} {\bf 83} (2023), no.~12 1122,
  [\href{http://xxx.lanl.gov/abs/2305.0171}{{\tt arXiv:2305.0171}}].

\bibitem{Ackerman:2008kmp}
L.~Ackerman, M.~R. Buckley, S.~M. Carroll, and M.~Kamionkowski, {\it {Dark
  Matter and Dark Radiation}},  {\em Phys. Rev. D} {\bf 79} (2009) 023519,
  [\href{http://xxx.lanl.gov/abs/0810.5126}{{\tt arXiv:0810.5126}}].

\bibitem{Feng:2008mu}
J.~L. Feng, H.~Tu, and H.-B. Yu, {\it {Thermal Relics in Hidden Sectors}},
  {\em JCAP} {\bf 10} (2008) 043,
  [\href{http://xxx.lanl.gov/abs/0808.2318}{{\tt arXiv:0808.2318}}].

\bibitem{Chu:2011be}
X.~Chu, T.~Hambye, and M.~H.~G. Tytgat, {\it {The Four Basic Ways of Creating
  Dark Matter Through a Portal}},  {\em JCAP} {\bf 05} (2012) 034,
  [\href{http://xxx.lanl.gov/abs/1112.0493}{{\tt arXiv:1112.0493}}].

\bibitem{Fabbrichesi:2020wbt}
M.~Fabbrichesi, E.~Gabrielli, and G.~Lanfranchi, {\it {The Dark Photon}},
  \href{http://xxx.lanl.gov/abs/2005.0151}{{\tt arXiv:2005.0151}}.

\bibitem{Pospelov:2007mp}
M.~Pospelov, A.~Ritz, and M.~B. Voloshin, {\it {Secluded WIMP Dark Matter}},
  {\em Phys. Lett. B} {\bf 662} (2008) 53--61,
  [\href{http://xxx.lanl.gov/abs/0711.4866}{{\tt arXiv:0711.4866}}].

\bibitem{Bernal:2015ova}
N.~Bernal, X.~Chu, C.~Garcia-Cely, T.~Hambye, and B.~Zaldivar, {\it {Production
  Regimes for Self-Interacting Dark Matter}},  {\em JCAP} {\bf 03} (2016) 018,
  [\href{http://xxx.lanl.gov/abs/1510.0806}{{\tt arXiv:1510.0806}}].

\bibitem{Slatyer:2015jla}
T.~R. Slatyer, {\it {Indirect dark matter signatures in the cosmic dark ages.
  I. Generalizing the bound on s-wave dark matter annihilation from Planck
  results}},  {\em Phys. Rev. D} {\bf 93} (2016), no.~2 023527,
  [\href{http://xxx.lanl.gov/abs/1506.0381}{{\tt arXiv:1506.0381}}].

\bibitem{DAgnolo:2015ujb}
R.~T. D'Agnolo and J.~T. Ruderman, {\it {Light Dark Matter from Forbidden
  Channels}},  {\em Phys. Rev. Lett.} {\bf 115} (2015), no.~6 061301,
  [\href{http://xxx.lanl.gov/abs/1505.0710}{{\tt arXiv:1505.0710}}].

\bibitem{DelNobile:2015uua}
E.~Del~Nobile, M.~Kaplinghat, and H.-B. Yu, {\it {Direct Detection Signatures
  of Self-Interacting Dark Matter with a Light Mediator}},  {\em JCAP} {\bf 10}
  (2015) 055, [\href{http://xxx.lanl.gov/abs/1507.0400}{{\tt
  arXiv:1507.0400}}].

\bibitem{Evans:2017kti}
J.~A. Evans, S.~Gori, and J.~Shelton, {\it {Looking for the WIMP Next Door}},
  {\em JHEP} {\bf 02} (2018) 100,
  [\href{http://xxx.lanl.gov/abs/1712.0397}{{\tt arXiv:1712.0397}}].

\bibitem{Hambye:2019dwd}
T.~Hambye, M.~H.~G. Tytgat, J.~Vandecasteele, and L.~Vanderheyden, {\it {Dark
  matter from dark photons: a taxonomy of dark matter production}},  {\em Phys.
  Rev. D} {\bf 100} (2019), no.~9 095018,
  [\href{http://xxx.lanl.gov/abs/1908.0986}{{\tt arXiv:1908.0986}}].

\bibitem{Fernandez:2021iti}
N.~Fernandez, Y.~Kahn, and J.~Shelton, {\it {Freeze-in, glaciation, and UV
  sensitivity from light mediators}},  {\em JHEP} {\bf 07} (2022) 044,
  [\href{http://xxx.lanl.gov/abs/2111.1370}{{\tt arXiv:2111.1370}}].

\bibitem{Bhattiprolu:2023akk}
P.~N. Bhattiprolu, R.~McGehee, and A.~Pierce, {\it {Dark sink enhances the
  direct detection of freeze-in dark matter}},  {\em Phys. Rev. D} {\bf 110}
  (2024), no.~3 L031702, [\href{http://xxx.lanl.gov/abs/2312.1415}{{\tt
  arXiv:2312.1415}}].

\bibitem{Boddy:2024vgt}
K.~K. Boddy, K.~Freese, G.~Montefalcone, and B.~Shams Es~Haghi, {\it {Minimal
  dark matter freeze-in with low reheating temperatures and implications for
  direct detection}},  {\em Phys. Rev. D} {\bf 111} (2025), no.~6 063537,
  [\href{http://xxx.lanl.gov/abs/2405.0622}{{\tt arXiv:2405.0622}}].

\bibitem{Berryman:2019dme}
J.~M. Berryman, A.~de~Gouvea, P.~J. Fox, B.~J. Kayser, K.~J. Kelly, and J.~L.
  Raaf, {\it {Searches for Decays of New Particles in the DUNE Multi-Purpose
  Near Detector}},  {\em JHEP} {\bf 02} (2020) 174,
  [\href{http://xxx.lanl.gov/abs/1912.0762}{{\tt arXiv:1912.0762}}].

\bibitem{DUNE:2020ypp}
{\bf DUNE} Collaboration, B.~Abi et~al., {\it {Deep Underground Neutrino
  Experiment (DUNE), Far Detector Technical Design Report, Volume II: DUNE
  Physics}},  \href{http://xxx.lanl.gov/abs/2002.0300}{{\tt arXiv:2002.0300}}.

\bibitem{SHiP:2021nfo}
{\bf SHiP} Collaboration, C.~Ahdida et~al., {\it {The SHiP experiment at the
  proposed CERN SPS Beam Dump Facility}},  {\em Eur. Phys. J. C} {\bf 82}
  (2022), no.~5 486, [\href{http://xxx.lanl.gov/abs/2112.0148}{{\tt
  arXiv:2112.0148}}].

\bibitem{Alekhin:2015byh}
S.~Alekhin et~al., {\it {A facility to Search for Hidden Particles at the CERN
  SPS: the SHiP physics case}},  {\em Rept. Prog. Phys.} {\bf 79} (2016),
  no.~12 124201, [\href{http://xxx.lanl.gov/abs/1504.0485}{{\tt
  arXiv:1504.0485}}].

\bibitem{Berlin:2018pwi}
A.~Berlin, S.~Gori, P.~Schuster, and N.~Toro, {\it {Dark Sectors at the
  Fermilab SeaQuest Experiment}},  {\em Phys. Rev. D} {\bf 98} (2018), no.~3
  035011, [\href{http://xxx.lanl.gov/abs/1804.0066}{{\tt arXiv:1804.0066}}].

\bibitem{Apyan:2022tsd}
A.~Apyan et~al., {\it {DarkQuest: A dark sector upgrade to SpinQuest at the 120
  GeV Fermilab Main Injector}},  in {\em {Snowmass 2021}}, 3, 2022.
\newblock \href{http://xxx.lanl.gov/abs/2203.0832}{{\tt arXiv:2203.0832}}.

\bibitem{Stueckelberg:1938hvi}
E.~C.~G. Stueckelberg, {\it {Interaction energy in electrodynamics and in the
  field theory of nuclear forces}},  {\em Helv. Phys. Acta} {\bf 11} (1938)
  225--244.

\bibitem{Redi:2022zkt}
M.~Redi and A.~Tesi, {\it {Dark photon Dark Matter without Stueckelberg mass}},
   {\em JHEP} {\bf 10} (2022) 167,
  [\href{http://xxx.lanl.gov/abs/2204.1427}{{\tt arXiv:2204.1427}}].

\bibitem{An:2013yfc}
H.~An, M.~Pospelov, and J.~Pradler, {\it {New stellar constraints on dark
  photons}},  {\em Phys. Lett. B} {\bf 725} (2013) 190--195,
  [\href{http://xxx.lanl.gov/abs/1302.3884}{{\tt arXiv:1302.3884}}].

\bibitem{Redondo:2013lna}
J.~Redondo and G.~Raffelt, {\it {Solar constraints on hidden photons
  re-visited}},  {\em JCAP} {\bf 08} (2013) 034,
  [\href{http://xxx.lanl.gov/abs/1305.2920}{{\tt arXiv:1305.2920}}].

\bibitem{Knapen:2017xzo}
S.~Knapen, T.~Lin, and K.~M. Zurek, {\it {Light Dark Matter: Models and
  Constraints}},  {\em Phys. Rev. D} {\bf 96} (2017), no.~11 115021,
  [\href{http://xxx.lanl.gov/abs/1709.0788}{{\tt arXiv:1709.0788}}].

\bibitem{Hardy:2016kme}
E.~Hardy and R.~Lasenby, {\it {Stellar cooling bounds on new light particles:
  plasma mixing effects}},  {\em JHEP} {\bf 02} (2017) 033,
  [\href{http://xxx.lanl.gov/abs/1611.0585}{{\tt arXiv:1611.0585}}].

\bibitem{Heeba:2019jho}
S.~Heeba and F.~Kahlhoefer, {\it {Probing the freeze-in mechanism in dark
  matter models with U(1)' gauge extensions}},  {\em Phys. Rev. D} {\bf 101}
  (2020), no.~3 035043, [\href{http://xxx.lanl.gov/abs/1908.0983}{{\tt
  arXiv:1908.0983}}].

\bibitem{Giudice:2003jh}
G.~F. Giudice, A.~Notari, M.~Raidal, A.~Riotto, and A.~Strumia, {\it {Towards a
  complete theory of thermal leptogenesis in the SM and MSSM}},  {\em Nucl.
  Phys. B} {\bf 685} (2004) 89--149,
  [\href{http://xxx.lanl.gov/abs/hep-ph/0310123}{{\tt hep-ph/0310123}}].

\bibitem{Gondolo:1990dk}
P.~Gondolo and G.~Gelmini, {\it {Cosmic abundances of stable particles:
  Improved analysis}},  {\em Nucl. Phys. B} {\bf 360} (1991) 145--179.

\bibitem{Evans:2019vxr}
J.~A. Evans, C.~Gaidau, and J.~Shelton, {\it {Leak-in Dark Matter}},  {\em
  JHEP} {\bf 01} (2020) 032, [\href{http://xxx.lanl.gov/abs/1909.0467}{{\tt
  arXiv:1909.0467}}].

\bibitem{Planck:2018vyg}
{\bf Planck} Collaboration, N.~Aghanim et~al., {\it {Planck 2018 results. VI.
  Cosmological parameters}},  {\em Astron. Astrophys.} {\bf 641} (2020) A6,
  [\href{http://xxx.lanl.gov/abs/1807.0620}{{\tt arXiv:1807.0620}}]. [Erratum:
  Astron.Astrophys. 652, C4 (2021)].

\bibitem{Dvorkin:2022bsc}
C.~Dvorkin et~al., {\it {Dark Matter Physics from the CMB-S4 Experiment}},  in
  {\em {Snowmass 2021}}, 3, 2022.
\newblock \href{http://xxx.lanl.gov/abs/2203.0706}{{\tt arXiv:2203.0706}}.

\bibitem{Berger:2018xyd}
J.~Berger, D.~Croon, S.~El~Hedri, K.~Jedamzik, A.~Perko, and D.~G.~E. Walker,
  {\it {Dark matter amnesia in out-of-equilibrium scenarios}},  {\em JCAP} {\bf
  02} (2019) 051, [\href{http://xxx.lanl.gov/abs/1812.0879}{{\tt
  arXiv:1812.0879}}].

\bibitem{Arvanitaki:2021qlj}
A.~Arvanitaki, S.~Dimopoulos, M.~Galanis, D.~Racco, O.~Simon, and J.~O.
  Thompson, {\it {Dark QED from inflation}},  {\em JHEP} {\bf 11} (2021) 106,
  [\href{http://xxx.lanl.gov/abs/2108.0482}{{\tt arXiv:2108.0482}}].

\bibitem{Batell:2022dpx}
B.~Batell, N.~Blinov, C.~Hearty, and R.~McGehee, {\it {Exploring Dark Sector
  Portals with High Intensity Experiments}},  in {\em {Snowmass 2021}}, 7,
  2022.
\newblock \href{http://xxx.lanl.gov/abs/2207.0690}{{\tt arXiv:2207.0690}}.

\bibitem{Chang:2016ntp}
J.~H. Chang, R.~Essig, and S.~D. McDermott, {\it {Revisiting Supernova 1987A
  Constraints on Dark Photons}},  {\em JHEP} {\bf 01} (2017) 107,
  [\href{http://xxx.lanl.gov/abs/1611.0386}{{\tt arXiv:1611.0386}}].

\bibitem{Fermi-LAT:2015att}
{\bf Fermi-LAT} Collaboration, M.~Ackermann et~al., {\it {Searching for Dark
  Matter Annihilation from Milky Way Dwarf Spheroidal Galaxies with Six Years
  of Fermi Large Area Telescope Data}},  {\em Phys. Rev. Lett.} {\bf 115}
  (2015), no.~23 231301, [\href{http://xxx.lanl.gov/abs/1503.0264}{{\tt
  arXiv:1503.0264}}].

\bibitem{Fermi-LAT:2016uux}
{\bf Fermi-LAT, DES} Collaboration, A.~Albert et~al., {\it {Searching for Dark
  Matter Annihilation in Recently Discovered Milky Way Satellites with
  Fermi-LAT}},  {\em Astrophys. J.} {\bf 834} (2017), no.~2 110,
  [\href{http://xxx.lanl.gov/abs/1611.0318}{{\tt arXiv:1611.0318}}].

\bibitem{AMS:2013fma}
{\bf AMS} Collaboration, M.~Aguilar et~al., {\it {First Result from the Alpha
  Magnetic Spectrometer on the International Space Station: Precision
  Measurement of the Positron Fraction in Primary Cosmic Rays of
  0.5\textendash{}350 GeV}},  {\em Phys. Rev. Lett.} {\bf 110} (2013) 141102.

\bibitem{Planck:2015fie}
{\bf Planck} Collaboration, P.~A.~R. Ade et~al., {\it {Planck 2015 results.
  XIII. Cosmological parameters}},  {\em Astron. Astrophys.} {\bf 594} (2016)
  A13, [\href{http://xxx.lanl.gov/abs/1502.0158}{{\tt arXiv:1502.0158}}].

\bibitem{Sommerfeld:1931qaf}
A.~Sommerfeld, {\it {\"Uber die Beugung und Bremsung der Elektronen}},  {\em
  Annalen Phys.} {\bf 403} (1931), no.~3 257--330.

\bibitem{Hisano:2002fk}
J.~Hisano, S.~Matsumoto, and M.~M. Nojiri, {\it {Unitarity and higher order
  corrections in neutralino dark matter annihilation into two photons}},  {\em
  Phys. Rev. D} {\bf 67} (2003) 075014,
  [\href{http://xxx.lanl.gov/abs/hep-ph/0212022}{{\tt hep-ph/0212022}}].

\bibitem{Hisano:2003ec}
J.~Hisano, S.~Matsumoto, and M.~M. Nojiri, {\it {Explosive dark matter
  annihilation}},  {\em Phys. Rev. Lett.} {\bf 92} (2004) 031303,
  [\href{http://xxx.lanl.gov/abs/hep-ph/0307216}{{\tt hep-ph/0307216}}].

\bibitem{Hisano:2004ds}
J.~Hisano, S.~Matsumoto, M.~M. Nojiri, and O.~Saito, {\it {Non-perturbative
  effect on dark matter annihilation and gamma ray signature from galactic
  center}},  {\em Phys. Rev. D} {\bf 71} (2005) 063528,
  [\href{http://xxx.lanl.gov/abs/hep-ph/0412403}{{\tt hep-ph/0412403}}].

\bibitem{Arkani-Hamed:2008hhe}
N.~Arkani-Hamed, D.~P. Finkbeiner, T.~R. Slatyer, and N.~Weiner, {\it {A Theory
  of Dark Matter}},  {\em Phys. Rev. D} {\bf 79} (2009) 015014,
  [\href{http://xxx.lanl.gov/abs/0810.0713}{{\tt arXiv:0810.0713}}].

\bibitem{Cirelli:2007xd}
M.~Cirelli, A.~Strumia, and M.~Tamburini, {\it {Cosmology and Astrophysics of
  Minimal Dark Matter}},  {\em Nucl. Phys. B} {\bf 787} (2007) 152--175,
  [\href{http://xxx.lanl.gov/abs/0706.4071}{{\tt arXiv:0706.4071}}].

\bibitem{Tulin:2013teo}
S.~Tulin, H.-B. Yu, and K.~M. Zurek, {\it {Beyond Collisionless Dark Matter:
  Particle Physics Dynamics for Dark Matter Halo Structure}},  {\em Phys. Rev.
  D} {\bf 87} (2013), no.~11 115007,
  [\href{http://xxx.lanl.gov/abs/1302.3898}{{\tt arXiv:1302.3898}}].

\bibitem{Cassel:2009wt}
S.~Cassel, {\it {Sommerfeld factor for arbitrary partial wave processes}},
  {\em J. Phys. G} {\bf 37} (2010) 105009,
  [\href{http://xxx.lanl.gov/abs/0903.5307}{{\tt arXiv:0903.5307}}].

\bibitem{Buschmann:2015awa}
M.~Buschmann, J.~Kopp, J.~Liu, and P.~A.~N. Machado, {\it {Lepton Jets from
  Radiating Dark Matter}},  {\em JHEP} {\bf 07} (2015) 045,
  [\href{http://xxx.lanl.gov/abs/1505.0745}{{\tt arXiv:1505.0745}}].

\bibitem{Madhavacheril:2013cna}
M.~S. Madhavacheril, N.~Sehgal, and T.~R. Slatyer, {\it {Current Dark Matter
  Annihilation Constraints from CMB and Low-Redshift Data}},  {\em Phys. Rev.
  D} {\bf 89} (2014) 103508, [\href{http://xxx.lanl.gov/abs/1310.3815}{{\tt
  arXiv:1310.3815}}].

\bibitem{Feng:2022inv}
J.~L. Feng et~al., {\it {The Forward Physics Facility at the High-Luminosity
  LHC}},  {\em J. Phys. G} {\bf 50} (2023), no.~3 030501,
  [\href{http://xxx.lanl.gov/abs/2203.0509}{{\tt arXiv:2203.0509}}].

\bibitem{Lewin:1995rx}
J.~D. Lewin and P.~F. Smith, {\it {Review of mathematics, numerical factors,
  and corrections for dark matter experiments based on elastic nuclear
  recoil}},  {\em Astropart. Phys.} {\bf 6} (1996) 87--112.

\bibitem{Freese:2012xd}
K.~Freese, M.~Lisanti, and C.~Savage, {\it {Colloquium: Annual modulation of
  dark matter}},  {\em Rev. Mod. Phys.} {\bf 85} (2013) 1561--1581,
  [\href{http://xxx.lanl.gov/abs/1209.3339}{{\tt arXiv:1209.3339}}].

\bibitem{Lin:2019uvt}
T.~Lin, {\it {Dark matter models and direct detection}},  {\em PoS} {\bf 333}
  (2019) 009, [\href{http://xxx.lanl.gov/abs/1904.0791}{{\tt
  arXiv:1904.0791}}].

\bibitem{Hambye:2018dpi}
T.~Hambye, M.~H.~G. Tytgat, J.~Vandecasteele, and L.~Vanderheyden, {\it {Dark
  matter direct detection is testing freeze-in}},  {\em Phys. Rev. D} {\bf 98}
  (2018), no.~7 075017, [\href{http://xxx.lanl.gov/abs/1807.0502}{{\tt
  arXiv:1807.0502}}].

\bibitem{CRESST:2024cpr}
{\bf CRESST} Collaboration, G.~Angloher et~al., {\it {First observation of
  single photons in a CRESST detector and new dark matter exclusion limits}},
  {\em Phys. Rev. D} {\bf 110} (2024), no.~8 083038,
  [\href{http://xxx.lanl.gov/abs/2405.0652}{{\tt arXiv:2405.0652}}].

\bibitem{CRESST:2019jnq}
{\bf CRESST} Collaboration, A.~H. Abdelhameed et~al., {\it {First results from
  the CRESST-III low-mass dark matter program}},  {\em Phys. Rev. D} {\bf 100}
  (2019), no.~10 102002, [\href{http://xxx.lanl.gov/abs/1904.0049}{{\tt
  arXiv:1904.0049}}].

\bibitem{DarkSide-50:2022qzh}
{\bf DarkSide-50} Collaboration, P.~Agnes et~al., {\it {Search for low-mass
  dark matter WIMPs with 12~ton-day exposure of DarkSide-50}},  {\em Phys. Rev.
  D} {\bf 107} (2023), no.~6 063001,
  [\href{http://xxx.lanl.gov/abs/2207.1196}{{\tt arXiv:2207.1196}}].

\bibitem{XENON:2024hup}
{\bf XENON} Collaboration, E.~Aprile et~al., {\it {First Search for Light Dark
  Matter in the Neutrino Fog with XENONnT}},
  \href{http://xxx.lanl.gov/abs/2409.1786}{{\tt arXiv:2409.1786}}.

\bibitem{LZCollaboration:2024lux}
{\bf LZ Collaboration} Collaboration, J.~Aalbers et~al., {\it {Dark Matter
  Search Results from 4.2 Tonne-Years of Exposure of the LUX-ZEPLIN (LZ)
  Experiment}},  \href{http://xxx.lanl.gov/abs/2410.1703}{{\tt
  arXiv:2410.1703}}.

\bibitem{OHare:2021utq}
C.~A.~J. O'Hare, {\it {New Definition of the Neutrino Floor for Direct Dark
  Matter Searches}},  {\em Phys. Rev. Lett.} {\bf 127} (2021), no.~25 251802,
  [\href{http://xxx.lanl.gov/abs/2109.0311}{{\tt arXiv:2109.0311}}].

\bibitem{Kamada:2024ntk}
A.~Kamada, T.~Kuwahara, S.~Matsumoto, Y.~Watanabe, and Y.~Watanabe, {\it
  {Mediator Decay through mixing with Degenerate Spectrum}},
  \href{http://xxx.lanl.gov/abs/2404.0679}{{\tt arXiv:2404.0679}}.

\bibitem{LoChiatto:2024guj}
P.~Lo~Chiatto and F.~Yu, {\it {Consistent Electroweak Phenomenology of a Nearly
  Degenerate $Z'$ Boson}},  \href{http://xxx.lanl.gov/abs/2405.0339}{{\tt
  arXiv:2405.0339}}.

\bibitem{DAMIC-M:2025luv}
{\bf DAMIC-M} Collaboration, K.~Aggarwal et~al., {\it {Probing Benchmark Models
  of Hidden-Sector Dark Matter with DAMIC-M}},
  \href{http://xxx.lanl.gov/abs/2503.1461}{{\tt arXiv:2503.1461}}.

\bibitem{Curtin:2023bcf}
D.~Curtin, Y.~Kahn, and R.~Nguyen, {\it {Dark photons from charged pion
  bremsstrahlung at proton beam experiments}},  {\em Phys. Rev. D} {\bf 108}
  (2023), no.~9 095039, [\href{http://xxx.lanl.gov/abs/2305.1930}{{\tt
  arXiv:2305.1930}}].

\bibitem{Kyselov:2024dmi}
Y.~Kyselov and M.~Ovchynnikov, {\it {Searches for long-lived dark photons at
  proton accelerator experiments}},  {\em Phys. Rev. D} {\bf 111} (2025), no.~1
  015030, [\href{http://xxx.lanl.gov/abs/2409.1109}{{\tt arXiv:2409.1109}}].

\bibitem{Zhou:2024aeu}
T.~Zhou, R.~Plestid, K.~J. Kelly, N.~Blinov, and P.~J. Fox, {\it {Long-lived
  vectors from electromagnetic cascades at SHiP}},  {\em JHEP} {\bf 02} (2025)
  107, [\href{http://xxx.lanl.gov/abs/2412.0188}{{\tt arXiv:2412.0188}}].

\bibitem{Falk:1978kf}
S.~W. Falk and D.~N. Schramm, {\it {Limits From Supernovae on Neutrino
  Radiative Lifetimes}},  {\em Phys. Lett. B} {\bf 79} (1978) 511.

\bibitem{Sung:2019xie}
A.~Sung, H.~Tu, and M.-R. Wu, {\it {New constraint from supernova explosions on
  light particles beyond the Standard Model}},  {\em Phys. Rev. D} {\bf 99}
  (2019), no.~12 121305, [\href{http://xxx.lanl.gov/abs/1903.0792}{{\tt
  arXiv:1903.0792}}].

\bibitem{Shin:2022ulh}
C.~S. Shin and S.~Yun, {\it {Dark gauge boson emission from supernova pions}},
  {\em Phys. Rev. D} {\bf 108} (2023), no.~5 055014,
  [\href{http://xxx.lanl.gov/abs/2211.1567}{{\tt arXiv:2211.1567}}].

\bibitem{Caputo:2022mah}
A.~Caputo, H.-T. Janka, G.~Raffelt, and E.~Vitagliano, {\it {Low-Energy
  Supernovae Severely Constrain Radiative Particle Decays}},  {\em Phys. Rev.
  Lett.} {\bf 128} (2022), no.~22 221103,
  [\href{http://xxx.lanl.gov/abs/2201.0989}{{\tt arXiv:2201.0989}}].

\bibitem{Tucker-Smith:2001myb}
D.~Tucker-Smith and N.~Weiner, {\it {Inelastic dark matter}},  {\em Phys. Rev.
  D} {\bf 64} (2001) 043502,
  [\href{http://xxx.lanl.gov/abs/hep-ph/0101138}{{\tt hep-ph/0101138}}].

\bibitem{CarrilloGonzalez:2021lxm}
M.~Carrillo~Gonz\'alez and N.~Toro, {\it {Cosmology and signals of light
  pseudo-Dirac dark matter}},  {\em JHEP} {\bf 04} (2022) 060,
  [\href{http://xxx.lanl.gov/abs/2108.1342}{{\tt arXiv:2108.1342}}].

\bibitem{Heeba:2023bik}
S.~Heeba, T.~Lin, and K.~Schutz, {\it {Inelastic freeze-in}},  {\em Phys. Rev.
  D} {\bf 108} (2023), no.~9 095016,
  [\href{http://xxx.lanl.gov/abs/2304.0607}{{\tt arXiv:2304.0607}}].

\end{thebibliography}\endgroup
%%%%%%%%%%%%%%%%%%%%%%%%%%%%%%%%%%%%%%%%%%%%%%%%%%%%%%%%%

\end{document}